\documentclass[a4paper,11pt]{article}
\usepackage{jheppub} 
\usepackage{lineno}
\usepackage[utf8]{inputenc}
\usepackage{url}
\usepackage{amsthm,amsmath,latexsym,amssymb,amsfonts,amscd}
\usepackage{tensor}
\usepackage{mleftright}
\usepackage{graphicx}
\usepackage{tikz}
\usepackage{tikz-feynman}
\usetikzlibrary{calc}
\usepackage{hyperref}
\usepackage{tabularx}

\tikzset{snake it/.style={decorate, decoration=snake}}

\title{\boldmath The 2-Dimensional Dual of $\phi^4$ in AdS$_3$}

\newcommand{\defeq}{\mathrel{:\mkern-0.25mu=}}
\newcommand{\Op}{\mathcal{O}}

\theoremstyle{plain}
\newtheorem{conjecture}{Conjecture}[section]

\author{Weichen Xiao ,}
\author{Ivo Sachs}
\affiliation{Arnold-Sommerfeld-Center for Theoretical Physics, Ludwig-Maximilians-Universität München,\\
Theresienstr. 37, 80333 Munich, Germany}

\emailAdd{w.xiao@lmu.de}

\abstract{We study the correlation functions of a conformally coupled $\phi^4$-interacting theory in AdS$_3$ and its dual CFT$_2$. The one-loop diagram is not expressible in terms of known transcendental functions, but is shown to be expressible as an infinite sum of previously well-studied tree-level diagrams, and we compute this sum using several number-theoretic conjectures. This enables us to extract recursively, the analytic expressions of anomalous dimensions of all dual double-trace operators. In the $s$-channel various consistency checks were performed against established bootstrap method, while our results in the $t$- and $u$-channel are not available in any previous literature to our knowledge. }

\begin{document}
\maketitle
\flushbottom
\section{Introduction} 

Anti-de Sitter space (AdS) and de Sitter space (dS) are two of the simplest spaces with non-vanishing curvature. Furthermore, the isometry group of $\text{AdS}_{d+1}$ is isomorphic to the global conformal group acting on its $d$-dimensional boundary. Consequently, in an AdS bulk with fixed metric background, any theory with correlators invariant under AdS isometries are automatically conformal invariant, when the positions of operator insertions are taken the boundary limit. No conjectural input is required in this construction, as one is simply considering a restriction of the same underlying theory \cite{Heemskerk:2009pn}. Given such bulk correlators, it is then possible to extract certain data of the dual CFT, namely the operator spectrum and OPE coefficients \cite{Witten:1998qj}. Within this framework, numerous studies have been carried out for quantum field theory in AdS at a variety of dimensions. An example is \cite{Paulos:2016fap}, where correlators in AdS were studied using conformal bootstrap techniques on the boundary, so that bounds on the flat-space amplitudes can be recovered by taking the AdS radius to its infinite limit. Some other examples of similar spirit include \cite{Aharony:2016dwx, DiPietro:2021sjt, Loparco:2026fki}.

As another concrete illustration of this idea, we consider a conformally coupled $\phi^4$ interacting scalar field in the bulk of $\text{AdS}$ and attempt to extract its dual conformal data corresponding to loop corrections. Bulk loops in Anti-de Sitter space have been considered for some time \cite{Bertan:2018khc,Bertan:2018afl,Carmi:2018qzm,Carmi:2019ocp,Cacciatori:2024zbe}. In \cite{Bertan:2018khc,Bertan:2018afl}, bulk one-loop diagrams for a conformally coupled scalar field with conformal dimensions $\Delta_- = 1$ and $\Delta_+ = 2$ were computed in $\text{AdS}_4$ and subsequently expanded in conformal cross ratios, enabling the extraction of anomalous dimensions in the dual perturbative CFT. A closed-form expression for these anomalous dimensions was later obtained in \cite{Heckelbacher:2022fbx}. In addition, the same concept was extended to the $\text{dS}_4/\text{CFT}_3$ in \cite{Heckelbacher:2020nue}. It is discovered in \cite{Sleight:2019hfp,Sleight:2021plv,DiPietro:2021sjt,Heckelbacher:2022hbq} that the cosmological in-in correlators in de Sitter space manifest as Witten diagrams involving a mixture of propagators with conformal dimension $\Delta_\pm$. As a result, such correlators may be computed using the techniques developed in the aforementioned works, providing an alternative to the Schwinger-Keldysh formalism. This observation has motivated further studies of cosmological correlators, including ~\cite{Chowdhury:2023arc,Nowinski:2025cvw}.

In this work, we study the $\text{AdS}_3$ case and compute the CFT data on the two-dimensional boundary, dual to the four-point one-loop diagrams of a conformally coupled scalar field with conformal dimension $\Delta = \tfrac{3}{2}$. We focus primarily on the $t$- and $u$-channels, where our analysis yields new results, while also considering the $s$-channel as a consistency check. The half-integer conformal dimension leads to non-integer powers in the Feynman integrals, rendering the corresponding computations effectively one-loop order more involved than their counterparts in $\text{AdS}_4$. As a consequence, we are unable to perform the conformal block expansion following the same approach as ~\cite{Bertan:2018khc,Bertan:2018afl}. Instead, we demonstrate a method in which loop-level amplitudes are expanded as an infinite series of tree-level diagrams with increasing conformal weights on their external legs, whose conformal block expansions have been studied extensively in ~\cite{Dolan:2001ih}. With the aid of two mathematical conjectures, we are able to resum this series and recover the analytic values of anomalous dimensions $\gamma_{n,l}$ of double trace operators $\phi\Box^n\partial^l\phi$ in the dual CFT.

In the $s$-channel, our direct bulk computation does not yield new results beyond the CFT data already known from \cite{Carmi:2018qzm}. Nevertheless, it provides a useful consistency check between bootstrap and perturbative calculations. In \cite{Carmi:2018qzm}, the spectral representation of the one-loop bubble diagram in AdS was obtained through a bootstrap approach, by imposing constraints from pole cancellation and asymptotic behavior of the spectral function. This allows one to explicitly extract the anomalous dimensions and OPE coefficients. On the other hand, by applying the aforementioned series expansion trick, we obtained an alternative representation of the same one-loop spectral function as an infinite sum of tree-level spectral functions. These tree-level spectral functions can be computed using the unitarity-cut method \cite{Meltzer:2019nbs}, which rewrites the contact 'cross' diagrams in terms of three-point functions. We then demonstrate that the two representations agree, giving rise to the same set of dual CFT data.

The novelty of our work lies in the $t$- and $u$-channel. Our series expansion trick for the one-loop diagram works regardless of whether we consider the $s$-, $t$- or $u$-channel. In contrast, the bootstrap method is unable to directly determine the $t$- and $u$-channel spectral functions in terms of the $s$-channel conformal partial waves due to the presence of nontrivial spin exchange in the double-trace operators. In principle, one may instead construct the $t$- and $u$-channel spectral functions in terms of their respective conformal partial waves and subsequently transform them to the $s$-channel basis using the conformal 6j symbols \cite{Liu:2018jhs} \cite{Meltzer:2019nbs}. This approach has been successfully applied in a number of settings, most notably in a recent study of the $O(N)$ model \cite{Dujava:2025php}. Unfortunately, we found its implementation in the present model — a single-component $\phi^4$ theory in $d=2$ with $\Delta=\tfrac{3}{2}$ — to be challenging in practice. It is indeed possible to write down the crossing kernels, following the approach of \cite{Dujava:2025php}[3.23]. However, it is challenging to explicitly extract anomalous dimensions due to the proliferation of hypergeometric functions in the resulting expressions. \footnote{Update at the time of JHEP review: \cite{Carmi:2026spv} has successfully overcome the difficulties associated with the 6j symbols and derived closed-form expressions for the anomalous dimensions and OPE coefficients. At $d=2$ and $\Delta=\frac{3}{2}$, their anomalous dimensions agree with our results. We strongly encourage interested readers to check out this paper.} In \cite{Aharony:2016dwx}, anomalous dimensions from the $t$- and $u$-channel have been computed by solving the crossing equation and also from Mellin amplitude. However, their discussions at $\mathcal{O}(1/N^4)$ were mostly limited to vanishing twist, i.e. $\gamma^{(2)}_{0,l}$. Therefore, we adopt a different strategy than all those mentioned above and compute the corresponding Witten diagrams directly. To the best of our knowledge, the $\mathcal{O}(\lambda^2)$ $t$- and $u$-channel anomalous dimensions $\gamma^{(2),t|u}_{n,l}$ and their recursive relation in $n$ and $l$ presented in this work are the first results of their kind. 

Last but not the least, this work builds a connection between quantum field theory in curved spacetime and two-dimensional conformal field theory, which is a dimension where many interesting examples of CFT have been studied extensively. Our study in scalar field with $\Delta=\frac{3}{2}$ sheds light on the other possible conformally coupled scaling dimension, $\Delta=\frac{1}{2}$, which we believe might be connected to the two dimensional Ising model and intend to explore in a future work.

The main result of the paper are anomalous dimensions $\gamma^{(2),t|u}_{n,l}$ that deforms the double trace operators $\mathcal{O}\Box^n\partial^l\mathcal{O}$ for all $n$ and even $l$, as a result of $t$- and $u$-channel one-loop corrections in the bulk. Some of the lowest order examples of these anomalous dimensions are shown in Table \ref{AnomDimt+u}.
\begin{table}[t]
\centering
\resizebox{\textwidth}{!}{$
\begin{array}{c|ccccc}
\gamma^{(2),t+u}_{n,l} & l=0 & l=2 & l=4 & l=6 & l=8 \\
\noalign{\vskip 0.6ex}
\hline
\noalign{\vskip 0.6ex}
n=0 &
-\dfrac{\lambda^2}{96\pi^2} &
-\dfrac{\lambda^2}{3360\pi^2} &
-\dfrac{\lambda^2}{15840\pi^2} &
-\dfrac{\lambda^2}{43680\pi^2} &
-\dfrac{\lambda^2}{93024\pi^2}
\\[1.2em]
n=1 &
-\dfrac{13\lambda^2}{3360\pi^2} &
-\dfrac{61\lambda^2}{221760\pi^2} &
-\dfrac{47\lambda^2}{617760\pi^2} &
-\dfrac{253\lambda^2}{8062080\pi^2} &
-\dfrac{397\lambda^2}{24961440\pi^2}
\\[1.2em]
n=2 &
-\dfrac{2059\lambda^2}{997920\pi^2} &
-\dfrac{4943\lambda^2}{21621600\pi^2} &
-\dfrac{104779\lambda^2}{1396755360\pi^2} &
-\dfrac{297547\lambda^2}{8761465440\pi^2} &
-\dfrac{676747\lambda^2}{37067738400\pi^2}
\\[1.2em]
n=3 &
-\dfrac{1873\lambda^2}{1441440\pi^2} &
-\dfrac{131779\lambda^2}{698377680\pi^2} &
-\dfrac{1496021\lambda^2}{21416915520\pi^2} &
-\dfrac{1239361\lambda^2}{36506106000\pi^2} &
-\dfrac{57960383\lambda^2}{3029445165600\pi^2}
\\[1.2em]
n=4 &
-\dfrac{10485469\lambda^2}{11639628000\pi^2} &
-\dfrac{58993787\lambda^2}{374796021600\pi^2} &
-\dfrac{460967861\lambda^2}{7228208988000\pi^2} &
-\dfrac{23672291549\lambda^2}{722017764468000\pi^2} &
-\dfrac{13867617787\lambda^2}{722017764468000\pi^2}
\end{array}
$}
\caption{Some lowest order $\gamma^{(2),t+u}_{n,l}$}
\label{AnomDimt+u}
\end{table}
A recursive relation has been found governing all $t$+$u$-channel anomalous dimensions as a function of $n$ and $l$, which due to its lengthiness we present in Appendix \ref{AppendixRecursiveRelation}. Given the initial values, this recursive relation enables us to extract $\gamma^{(2)}_{n,l}$ at any given order of $n$ and $l$ efficiently. In addition, for $n=0$ we found a closed form expression
\begin{equation}
  \begin{aligned}
    \gamma^{s+t+u}_{0,l}=-\frac{\lambda^2}{32\pi^2 (l+1)(2l+1)(2l+3)}-\frac{\lambda^2}{256\pi^2}\delta^{l0}.
  \end{aligned}
\end{equation}

The paper will be structured as follows: in Chapter 2 we introduce the preliminaries of a conformally coupled $\phi^4$ theory in $\text{AdS}_3$ and its dual. In Chapter 3 we compute the tree level 'cross' diagram which not only contributes to the leading order anomalous dimensions and OPE coefficients, but also serves as building block for the one-loop diagram. In Chapter 4 we demonstrate the trick of expressing the one-loop diagram as an infinite sum of cross diagrams. As a consequence, in the $s$-channel the spectral function of the loop diagram can be read off directly as a corresponding sum. We then perform a consistency check by comparing this result with the spectral function obtained via bootstrap methods in \cite{Carmi:2018qzm}. In Chapter 5 we show that the same technique does not extend to the $t$- and $u$-channel. Instead, we carry out a conformal block expansion for each individual cross diagram appearing in the series. Subsequently, a resummation of the series expansion is performed using two conjectures proposed in this work, enabling us to guess a recursive relation governing the exact values of the anomalous dimensions.

Mathematica notebooks that were used to produce results in this paper can be accessed in the repository \href{https://github.com/weichenxiao/The-2-Dimensional-Dual-of-phi4-in-AdS3}{here}.
\section{Preliminaries}
The $d\!+\!1$ dimensional (Euclidean)anti-de Sitter space ($\text{(E)AdS}_{d+1}$) is defined by an embedding in a $d\!+\!2$ dimensional Lorentzian space with metric
\begin{equation}
    \eta_{\text{Euc}}=\text{diag}(1,\underbrace{-1, \ldots, -1}_{d+1\text{ times}})\,,
\end{equation}
which is considered a fixed background, as in this paper we do not take into account any back reaction. The embedding takes the form
\begin{equation}
    X^2 = X_0^2-\sum\limits_i X_i^2 = a^2.
\end{equation}
We will work on the Poincar\'e patch of anti-de Sitter space, parameterized by
\begin{equation}
  \begin{aligned}
X_0=&\frac{1}{2}\frac{z^2+\vec{x}^2+a^2}{z} \\
X_{d+1}=&\frac{1}{2}\frac{z^2+\vec{x}^2-a^2}{z} \\
X_i=&\frac{a}{z}x_i\,,
 \end{aligned}
\end{equation}
so that $\eta_{\text{Euc}}$ induces a hyperbolic metric
\begin{equation}
g_\text{AdS}= -\frac{1}{z^2}(dx_i^2 + dz^2).
\end{equation}
We study a simple model of $\phi^4$ interacting scalar field in $\text{AdS}_3$, governed by the following action
\begin{equation}
S=\int_{\text{AdS}_3} \text{d}^3x \sqrt{|g_\text{AdS}|} \left( \frac{1}{2}(\partial\phi)^2+\frac{1}{2}m^2\phi^2+\xi R\phi^2+\frac{\lambda}{4!}\phi^4 \right).
\end{equation}
Furthermore, we assume the scalar field to be conformally coupled, so that the mass term cancels with the curvature term for $\xi=\frac{d-1}{4d}$, leaving the Lagrangian analogous to massless scalar theory in flat space. Hence it is required that
\begin{equation}
  \begin{aligned}
m^2=-\frac{d^2-1}{4}a^2.
\label{CoupledMass}
 \end{aligned}
\end{equation}
Since the AdS background is fixed, any theory in such background has to be invariant under the isometry group of AdS. On the other hand, an isometry group in $\text{AdS}_{d+1}$ is isomorphic to a conformal group on the boundary $\text{CFT}_d$. Therefore the bulk theory automatically becomes a conformal field theory when the boundary limit is taken. A massive scalar field in the bulk is dual to an operator on the boundary with conformal dimension \cite{Aharony:1999ti}
\begin{equation}
  \begin{aligned}
\Delta=\frac{d}{2} \pm \sqrt{\frac{d^2}{4}+\frac{m^2}{a^2}}\,.
\label{ConformalDimensionspm}
 \end{aligned}
\end{equation}
For concreteness we will consider the + sign in this work. The minus sign does also have an interesting CFT dual that we will will consider in a forthcoming work. Combining it with (\ref{CoupledMass}) yields
\begin{equation}
  \begin{aligned}
\Delta=\frac{3}{2}.
 \end{aligned}
\end{equation}
The bulk to bulk propagator of such a scalar field then takes the form (e.g. \cite{Witten:1998qj})
\begin{equation}
\begin{aligned}
    \Lambda(x,y,\Delta=\frac{3}{2})&=\frac{a\sqrt{K_{xy}}}{4\sqrt{2}\pi}(\frac{1}{\sqrt{1-K_{xy}}}-\frac{1}{\sqrt{1+K_{xy}}}) \\
    &=\frac{a}{4\sqrt{2}\pi}\left( \sqrt{\frac{2zw}{(x^i-y^i)^2+(z-w)^2}} - \sqrt{\frac{2zw}{(x^i-y^i)^2+(z+w)^2}} \right)\,,
\end{aligned}
\label{BulktoBulkPropagator}
\end{equation}
where $K_{xy}$ is an isometry invariant quantity defined by
\begin{equation}
    K_{xy}=\frac{2zw}{(x^i-y^i)^2+z^2+w^2},
\end{equation}
that relates to the geodesic distance $\rho_{xy}$ through  $\frac{1}{K_{xy}}=\cosh{a\rho_{xy}}$.
The presence of the square root in \eqref{BulktoBulkPropagator} presents a significant complication for the computation of Feynman integrals compared to even dimensional anti-de Sitter spaces where the propagator comes with integer powers for conformally coupled fields.

With the point $(y,w)$ in (\ref{BulktoBulkPropagator}) taken to the boundary, the bulk to boundary propagators reduces to
\begin{equation}
  \begin{aligned}
\overline{\Lambda}(x,y,\Delta)=&\frac{a}{4\sqrt{2}\pi}2^{\frac{3}{2}-\Delta} w \overline{K}^\Delta \\
=&\frac{a}{2 \pi} \left(\frac{w z}{(x-y)^2+z^2}\right)^\Delta.
 \end{aligned}
 \label{BulktoBoundaryPropagator}
\end{equation}
Similarly, all positions of operator insertion in a four point function can be taken to the boundary, resulting in a conformal four point correlator \cite{Maldacena:1997re,Klebanov:1999tb}. 

In conformal field theory, such correlator can be written as
\begin{equation}
  \begin{aligned}
\langle\Op_\Delta(\vec{x}_1)\Op_\Delta(\vec{x}_2)\Op_\Delta(\vec{x}_3)\Op_\Delta(\vec{x}_4)\rangle =  \frac{N_\phi^2}{r_{12}^{2\Delta}r_{34}^{2\Delta}}\sum_{n=0,l=0,l\text{ even}}^\infty\mathcal{A}_{n,l}\mathcal{G}_{n,l}\,,
  \end{aligned}
\end{equation}
in which $\mathcal{G}_{n,l}=\mathcal{G}(\Delta(n,l),l)$ is the conformal block \cite{Dolan:2000ut}, defined in Appendix \ref{AppendixCBDef}, which forms a basis of conformal correlators. $\mathcal{A}_{n,l}=\mathcal{A}(\Delta(n,l),l)$ is the OPE coefficient squared, and $N_\phi$ is the normalization of bulk to boundary propagator that is factored out for consistency with \cite{Bertan:2018afl} \cite{Heckelbacher:2022fbx}. In our setting of $d=2$ and $\Delta=\frac{3}{2}$, $N_\phi=\frac{a}{2\pi}$. Due to crossing symmetry the conformal block expansion can be equivalently done in the $s$, $t$, or $u$-channel. Here we choose the $s$-channel which corresponds to the OPE for small $r_{12}$ and $r_{34}$.

For a free scalar field theory in AdS, its dual correlator is that of a generalized free field \cite{Heemskerk:2009pn,Greenberg:1961mr} for which the composite, or double trace operators of the form $\mathcal{O}_\Delta\Box^n\partial^l\mathcal{O}_\Delta$ carry conformal dimension
\begin{equation}
  \begin{aligned}
    \Delta^{\text{free}}(n,l)=2\Delta+2n+l.
  \end{aligned}
\end{equation}
The four point correlator of a free theory then admits the following OPE
\begin{equation}  
  \begin{aligned}
&\langle\Op_\Delta(\vec{x}_1)\Op_\Delta(\vec{x}_2)\Op_\Delta(\vec{x}_3)\Op_\Delta(\vec{x}_4)\rangle_{\text{disc}} \\
=& \sum_{n=0,l=0,l\text{ even}}^\infty \mathcal{A}(\Delta^{\text{free}}(n,l),l) \mathcal{G}(\Delta^{\text{free}}(n,l),l) \\
\defeq& \sum_{n=0,l=0,l\text{ even}}^\infty A_{n,l}G_{n,l}\,.
\label{CBExpansionFreeTheory}
  \end{aligned}
\end{equation}
The left hand side can be expressed as a disconnected AdS field theory 4-point function \cite{Bertan:2018afl}:
\begin{equation}
  \begin{aligned}
&\langle\Op_\Delta(\vec{x}_1)\Op_\Delta(\vec{x}_2)\Op_\Delta(\vec{x}_3)\Op_\Delta(\vec{x}_4)\rangle_{\text{disc}} \\
=&\quad\diagramtwob \\
=&  \Lambda(x_1,x_3)\Lambda(x_2,x_4)+\Lambda(x_1,x_2)\Lambda(x_3,x_4)+\Lambda(x_1,x_4)\Lambda(x_2,x_3) \\
=&\frac{N_\phi^2}{(r_{12}r_{34})^{2\Delta}} \left( 1+  u^\Delta  + \frac{u^\Delta}{v^\Delta} \right) \\
=&\frac{N_\phi^2}{(r_{12}r_{34})^{2\Delta}} \left( 1+  u^\Delta \left( 2+\sum_{n=1}^\infty \frac{\Gamma(\Delta+n)}{\Gamma(\Delta)\Gamma(n+1)} \right)(1-v)^n \right)\,,
\label{FreeTheoryCorrelator}
  \end{aligned}
\end{equation}
where $r_{ij}=|\vec{x}_i-\vec{x}_j|$ and $N_\phi=\frac{a}{2\pi}$ originated from (\ref{BulktoBoundaryPropagator}). The conformal cross ratios $u$ and $v$ are defined by
\begin{equation}
  \begin{aligned}
u=\frac{r_{12}^2 r_{34}^2}{r_{13}^2 r_{24}^2} \quad v=\frac{r_{14}^2 r_{23}^2}{r_{13}^2 r_{24}^2}.
 \end{aligned}
\end{equation}
At higher loop order, $\Delta(n,l)$ deviates from its generalized free field value, obtaining corrections of the form
\begin{equation}\label{e:Deltacorrect}
    \Delta_{(n,l)}\to\Delta_{(n,l)}+\sum\limits_{p=0}^\infty\gamma_{n,l}^{(p)}(\Delta)\,.
\end{equation}
This results in $\mathcal{A}_{n,l}$ and $\mathcal{G}_{n,l}$ also being perturbatively corrected as follows,
\begin{equation}
\begin{aligned}
	\mathcal{A}_{n,l}(\Delta)=&A_{n,l}(\Delta)+(\gamma^{(1)}_{n,l}(\Delta)+\gamma^{(2)}_{n,l}(\Delta))A^{(1)}_{n,l}+\frac12(\gamma^{(1)}_{n,l}(\Delta))^2A^{(2)}_{n,l}+ \cdots \\
\nonumber	\mathcal{G}_{n,l}=&G_{n,l}+(\gamma^{(1)}_{n,l}(\Delta)+\gamma^{(2)}_{n,l}(\Delta))\underbrace{\left.\frac{\partial G_{n,l}}{\partial\Delta}\right\vert_{\Delta(n,l)}}_{G'_{n,l}}
	+\frac12(\gamma^{(1)}_{n,l}(\Delta))^2\underbrace{\left.\frac{\partial^2G_{n,l}}{\partial\Delta^2}\right\vert_{\Delta(n,l)}}_{G''_{n,l}}+ \cdots\,,
\end{aligned}
\end{equation}
so that 
\begin{equation}
\begin{aligned}
\mathcal{A}_{n,l}\mathcal{G}_{n,l}=&A_{n,l}G_{n,l}+\gamma^{(1)}_{n,l}(\Delta)\left(A_{n,l}G'_{n,l}+A^{(1)}_{n,l}G_{n,l}\right)\cr
	&+\frac{1}{2}(\gamma^{(1)}_{n,l}(\Delta))^2\left(A_{n,l}G''_{n,l}+A^{(2)}_{n,l}G_{n,l}+2A^{(1)}_{n,l}G'_{n,l}\right)\cr
	&+\gamma^{(2)}_{n,l}(\Delta)\left(A_{n,l}G'_{n,l}+A^{(1)}_{n,l}G_{n,l}\right)+\mathcal{O}(\lambda^3)\,.
	\label{eq:CBExp}
\end{aligned}
\end{equation}
The goal of this paper is then to extract $\gamma^{(1)}_{n,l}$, $\gamma^{(2)}_{n,l}$, $A^{(1)}_{n,l}$ and $A^{(2)}_{n,l}$, which characterizes the CFT that is dual to the bulk scalar field at each perturbative order.

\section{Cross Diagram}
To lowest order in the perturbation, the bulk correction is given by the four point function, the cross diagram $\mathcal{I}_4$, as expressed in (\ref{I4}) and Figure \ref{fig:I4}.
\begin{equation}
  \begin{aligned}
\mathcal{I}_4=&\int \text{d}^{d+1}x \sqrt{|g_x|} \Lambda(x,x_1,\Delta_1) \Lambda(x,x_2,\Delta_2) \Lambda(x,x_3,\Delta_3) \Lambda(x,x_4,\Delta_4)
\label{I4}
 \end{aligned}
\end{equation}
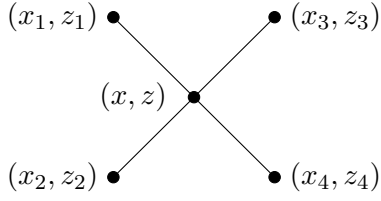
\begin{figure}[t]
\begin{tikzpicture}
  \tikzfeynmanset{
    every vertex={black, dot},
  }
  \begin{feynman}
    \vertex (a);
    \vertex [left=0.8cm of a] (atext) {\((x,z)\)};
    \vertex [above left=of a] (f1);
    \vertex [left=0.8cm of f1] (f1text) {\((x_1,z_1)\)};
    \vertex [below left=of a] (f2);
    \vertex [left=0.8cm of f2] (f2text) {\((x_2,z_2)\)};
    \vertex [above right=of a] (f3);
    \vertex [right=0.8cm of f3] (f3text) {\((x_3,z_3)\)};
    \vertex [below right=of a] (f4);
    \vertex [right=0.8cm of f4] (f4text) {\((x_4,z_4)\)};

    \diagram* {
      (a) -- [plain] (f1),
      (a) -- [plain] (f2),
      (a) -- [plain] (f3),
      (a) -- [plain] (f4),
    };
  \end{feynman}
\end{tikzpicture}
\centering
\caption{The cross diagram \(\mathcal{I}_4\)}
\label{fig:I4}
\end{figure}
The method of calculating such integrals  is already well-developed in previous works \cite{Dolan:2000uw} and \cite{Dolan:2000ut} for general dimensions. Here we do a quick recap of their first few steps in order to fix our normalization relative to the references. Inserting the bulk to boundary propagator (\ref{BulktoBoundaryPropagator}), the cross diagram then reads
\begin{equation}
  \begin{aligned}
\mathcal{I}_4=&\prod_{i=1}^4 {z_i}^{\Delta_i} \int \text{d}^{d+1}y \frac{a^{4-d-1}}{16 \pi^4} z^{\Delta_1+\Delta_2+\Delta_3+\Delta_4-d-1} \prod_{i=1}^4 (\frac{1}{(x-x_i)^2+z^2})^{\Delta_i} \\
 \end{aligned}
\end{equation}
Making use of the Schwinger parametrization and defining $C=\frac{a^{4-d-1}}{16 \pi^4} \prod_{i=1}^4 {z_i}^{\Delta_i}$, $\Lambda=\sum_i \lambda_i$ and $\mu=\frac{1}{2}\sum_{i=1}^4\Delta_i$,
\begin{equation}
  \begin{aligned}
\mathcal{I}_4=&C \int^\infty_0 \text{d}z \frac{z^{2\mu-d-1}}{\prod_{i=1}^4\Gamma(\Delta_i)} \int^\infty_0 \text{d}\lambda_1 \text{d}\lambda_2 \text{d}\lambda_3 \text{d}\lambda_4 \\
&\prod_{i=1}^4\lambda_i^{\Delta_i} \pi^{\frac{1}{2}d} \Lambda^{-\frac{1}{2}d} e^{-\frac{1}{\Lambda}(\sum_{i<j}^4 \lambda_i\lambda_j r_{ij}^2)} e^{-\Lambda z^2}.
 \end{aligned}
\end{equation}
Substituting $t=\Lambda z^2$,
\begin{equation}
  \begin{aligned}
\mathcal{I}_4=&C \int^\infty_0 \text{d}t \text{d}\lambda_1 \text{d}\lambda_2 \text{d}\lambda_3 \text{d}\lambda_4\frac{t^{\mu-\frac{1}{2}d-1}}{2\prod_{i=1}^4\Gamma(\Delta_i) \Lambda^{\mu-\frac{1}{2}d}} \\
&\prod_{i=1}^4\lambda_i^{\Delta_i} \pi^{\frac{1}{2}d} \Lambda^{-\frac{1}{2}d} e^{-\frac{1}{\Lambda}(\sum_{i<j}^4 \lambda_i\lambda_j r_{ij}^2)} e^{-t}.
 \end{aligned}
\end{equation}
By the definition of Gamma function, this is equivalent to
\begin{equation}
  \begin{aligned}
\mathcal{I}_4=&C \frac{\pi^{\frac{1}{2}d} \Gamma(\mu-\frac{1}{2}d)}{2\prod_{i=1}^4\Gamma(\Delta_i)} \int^\infty_0 \text{d}\lambda_1 \text{d}\lambda_2 \text{d}\lambda_3 \text{d}\lambda_4 \frac{1}{\Lambda^{\mu}} \\
&\prod_{i=1}^4\lambda_i^{\Delta_i} e^{-\frac{1}{\Lambda}(\sum_{i<j}^4 \lambda_i\lambda_j r_{ij}^2)}.
 \end{aligned}
\end{equation}
Defining $I_4$ to be the same as in \cite[B.5]{Dolan:2000uw}, we have the following normalization
\begin{equation}
    \mathcal{I}_4=\frac{\pi^\frac{d}{2}}{\prod_{i=1}^4\Gamma(\Delta_i)}\left(\frac{a}{2\pi}\right)^4 \frac{\Gamma(\frac{\Delta_1+\Delta_2+\Delta_3+\Delta_4}{2}-\frac{d}{2})}{2}I_4
\label{cross}
\end{equation}
and the explicit expression of $I_4$ is given by \cite[B.11]{Dolan:2000uw}.

\section{Loop diagrams}
The main focus of this paper is to compute  anomalous dimensions and OPE coefficients due to the 1-loop corrections to the four point correlator as shown in (\ref{ThreeChannelsOfOneLoopCorrection}). 
\begin{equation}
\begin{aligned}
    W^{\rangle\!\circ\!\langle}=& \frac{\lambda^2}{2}(\mathcal{K}_4^s+\mathcal{K}_4^t+\mathcal{K}_4^u)\\
    =&\frac{\lambda^2}{2}\bigg( \underbrace{\diagramtwos}_{\text{$s$-channel}}+\underbrace{\diagramtwot}_{\text{$t$-channel}}+\underbrace{\diagramtwou}_{\text{$u$-channel}} \bigg),
\end{aligned}
\label{ThreeChannelsOfOneLoopCorrection}
\end{equation}
in which the factor $\frac{1}{2}$ came from the symmetry factor of the one-loop diagram. Unlike the cross diagram there is no general expression like \eqref{cross} for this integral for generic dimensions and conformal weights. For the $s$-channel diagram it is convenient to use a spectral representation. In \cite{Carmi:2018qzm} the spectral function $\tilde{B}(\nu)$ was then derived by a bootstrap approach resulting in a simple expression for $d<3$ \cite{Carmi:2018qzm}. For $d=3$,  $\sigma(\nu)$ is divergent and requires renormalization \cite{Sachs:2023eph}. 

For the $t$- and $u$-channel diagrams, the bootstrap of \cite{Carmi:2018qzm} does not directly apply. In principle, it is possible to extract the $t$-channel OPE from that of the  $s$-channel (e.g. \cite{Meltzer:2019nbs} \cite{Liu:2018jhs}). However, in practice we found it difficult to extract any explicit result, as it is challenging to evaluate the product of hypergeometric functions resulting from the conformal 6j symbol. We will comment more on this in Section \ref{sec:t-u}. Therefore, we will have to use a different approach in the following sections which we structure as follows: in Section \ref{SectionFishDiagram}. we discuss the fish diagram, as shown in Figure \ref{fig:J4}, which serves as a stepping stone for further computations of the full diagrams in all 3 channels. For the $s$-channel, we found two distinct methods to extract the dual CFT data: spectral function and conformal block expansion. The former method, demonstrated in Section \ref{SectionSchannelSpectralFunction}, serves as a crosscheck with well established results in \cite{Carmi:2018qzm}. We leave the latter method to Section \ref{SectionSchannelCBExpansion} where the consistency of the two methods will be shown in the $s$-channel. For the $t$- and $u$-channel, however, it is impossible to directly extract the spectral function and hence conformal block expansion is mandatory, which will be left to Section \ref{SectionTUchannelCBExpansion} as well.

\subsection{Fish diagram and tricks with conformal transformation}

\label{SectionFishDiagram}

\begin{figure}[t]
\begin{tikzpicture}
  \tikzfeynmanset{
    every vertex={black, dot},
  }
  \begin{feynman}
    \vertex (a);
    \vertex [left=0.7cm of a] (atext) {\((x,z)\)};
    \vertex [right=of a] (b);
    \vertex [right=0.8cm of b] (btext) {\((y,w)\)};
    \vertex [above right=of b] (f1);
    \vertex [right=0.8cm of f1] (f1text) {\((x_3,z_3)\)};
    \vertex [below right=of b] (f2);
    \vertex [right=0.8cm of f2] (f2text) {\((x_4,z_4)\)};

    \diagram* {
      (a)[edge label'=\(W^{-}\) ] -- [plain, half left] (b) -- [plain, half left] (a),
      (b) -- [plain] (f1),
      (b) -- [plain] (f2),
    };
  \end{feynman}
\end{tikzpicture}
\centering
\caption{The fish diagram \(\mathcal{J}_4\)}
\label{fig:J4}
\end{figure}
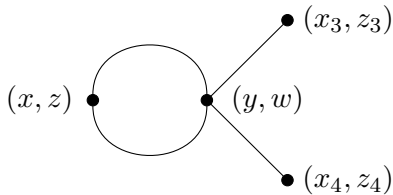

To start with, we first consider the 'fish' diagram $\mathcal{J}_4$ (as shown in Figure \ref{fig:J4}) to which we will eventually attach the two remaining external legs as in \cite{Bertan:2018afl}. It reads
\begin{equation}
\begin{aligned}
    \mathcal{J}_4=\int \text{d}^3y \sqrt{|g_y|} \Lambda(x,y)^2 \overline{\Lambda}(y,x_3) \overline{\Lambda}(y,x_4).
\end{aligned}
\end{equation}
Details of such loop integral are shown in Appendix \ref{AppendixJ4integraldetail}, giving the following result:
\begin{equation}
  \begin{aligned}
\mathcal{J}_4 =&\frac{a \lambda}{128\pi^3} (z_3 z_4)^\frac{3}{2} (\overline{K}_{x x_3}  \overline{K}_{x x_4})^\frac{3}{2} \left( \sqrt{1+\alpha^2}\text{EllipticE}[-\alpha^2] - (1+\alpha^2) \right),
 \end{aligned}
\end{equation}
where $\text{EllipticE}[x]$ is the complete elliptic integral of the second kind and the covariant quantity $\alpha=\frac{|x_3'''|}{z''}$ is related to the geodesic distance through
\begin{equation}
\frac{4 {z''}^2}{{x_3'''}^2+{z''}^2}=r_{34}^2 \overline{K}_{x x_3}  \overline{K}_{x x_4}=\frac{4}{\alpha^2+1}.
\label{recovercovariance}
\end{equation}
Proceeding as in 
$d=3$ would require evaluating integrals similar to those appearing in \cite[Appendix A]{Bertan:2018afl}, which we were unable to expand in terms of conformal invariants. Instead, a key observation of this paper is that the elliptic integral has a series expansion in $1+\alpha^2$, such that
\begin{equation}
  \begin{aligned}
\mathcal{J}_4=&\frac{a \lambda}{128\pi^3} (z_3 z_4)^\frac{3}{2} (\overline{K}_{x x_3}  \overline{K}_{x x_4})^\frac{3}{2} \sum_{k=0}^{\infty}\left( C_k (1+\alpha^2)^{-k} + D_k (1+\alpha^2)^{-k} \log(1+\alpha^2) \right) \\
=&\frac{a \lambda}{128\pi^3} (z_3 z_4)^\frac{3}{2} (\overline{K}_{x x_3}  \overline{K}_{x x_4})^\frac{3}{2} \sum_{k=0}^{\infty}\left( C_k (1+\alpha^2)^{-k} - D_k \frac{\text{d}}{\text{d}\epsilon}|_{\epsilon=0} (1+\alpha^2)^{-k-\epsilon} \right) \\
=&\frac{a \lambda}{128\pi^3} (z_3 z_4)^\frac{3}{2} (\overline{K}_{x x_3}  \overline{K}_{x x_4})^\frac{3}{2} \sum_{k=0}^{\infty} \left( C_k (\frac{r_{34}^2 \overline{K}_{x x_3}  \overline{K}_{x x_4}}{4})^{k} -  D_k \frac{\text{d}}{\text{d}\epsilon}|_{\epsilon=0} \left( (\frac{r_{34}^2 \overline{K}_{x x_3}  \overline{K}_{x x_4}}{4})^{k+\epsilon} \right) \right)
\label{EllipticExpansion}
  \end{aligned}
\end{equation}
with series coefficients
\begin{equation}
  \begin{aligned}
 C_k=-\frac{\left( \frac{1}{2} \right)_k \left( \frac{3}{2} \right)_k \left( 1+(2+6k+4k^2)\psi(\frac{1}{2}+k)-(2+6k+4k^2)\psi(1+k) \right)}{4(1+3k+2k^2)k!(1+k)!}\,,
 \end{aligned}
\end{equation}
and
\begin{equation}
  \begin{aligned}
 D_k=\frac{\left( \frac{1}{2} \right)_k \left( \frac{3}{2} \right)_k }{4k!(1+k)!}\,.
 \end{aligned}
\end{equation}
Then, recalling the bulk-to-boundary propagators from (\ref{BulktoBoundaryPropagator}), eqn. \eqref{EllipticExpansion} becomes
\begin{equation}
  \begin{aligned}
\mathcal{J}_4=&\frac{\lambda}{4\pi a} (z_3 z_4)^\frac{3}{2} \sum_{k=0}^{\infty} C_k (r_{34}^2)^k \overline{\Lambda}(x,x_3,\frac{3}{2}+k) \overline{\Lambda}(x,x_4,\frac{3}{2}+k)  \\
-&\frac{\lambda}{4\pi a} (z_3 z_4)^\frac{3}{2} \sum_{k=0}^{\infty}  D_k \frac{\text{d}}{\text{d}\epsilon}|_{\epsilon=0} (r_{34}^2)^{k+\epsilon} \overline{\Lambda}(x,x_3,\frac{3}{2}+k+\epsilon) \overline{\Lambda}(x,x_4,\frac{3}{2}+k+\epsilon)\,,
  \end{aligned}
\end{equation}
where the factor of $(\frac{1}{4})^{\frac{3}{2}+k+\epsilon}$ in (\ref{EllipticExpansion}) contributes to the $2^{\frac{3}{2}-\Delta}$ factor in the propagator (\ref{BulktoBoundaryPropagator}).
\begin{figure}[t]
\begin{tikzpicture}
  \tikzfeynmanset{
    every vertex={black, dot},
  }
  \begin{feynman}
    \vertex (a);
    \vertex [left=0.8cm of a] (atext) {\((x,z)\)};
    \vertex [above left=of a] (f1);
    \vertex [left=0.8cm of f1] (f1text) {\((x_1,z_1)\)};
    \vertex [below left=of a] (f2);
    \vertex [left=0.8cm of f2] (f2text) {\((x_2,z_2)\)};
    \vertex [right=of a] (b);
    \vertex [right=0.8cm of b] (btext) {\((y,w)\)};
    \vertex [above right=of b] (f3);
    \vertex [right=0.8cm of f3] (f3text) {\((x_3,z_3)\)};
    \vertex [below right=of b] (f4);
    \vertex [right=0.8cm of f4] (f4text) {\((x_4,z_4)\)};

    \diagram* {
      (a)[edge label'=\(W^{-}\) ] -- [plain, half left] (b) -- [plain, half left] (a),
      (a) -- [plain] (f1),
      (a) -- [plain] (f2),
      (b) -- [plain] (f3),
      (b) -- [plain] (f4),
    };
  \end{feynman}
\end{tikzpicture}
\centering
\caption{The full one-loop diagram \(\mathcal{K}_4\)}
\label{fig:K4}
\end{figure}
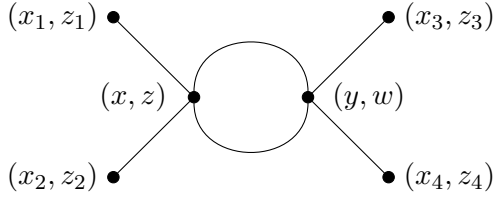
Attaching the two missing legs we then find
\begin{equation}
  \begin{aligned}
\mathcal{K}_4^s=&\int \text{d}^3x \frac{\lambda^2}{4\pi a} \sum_{k=0}^{\infty} C_k (r_{34}^2)^k \overline{\Lambda}(x,x_1,\frac{3}{2}) \overline{\Lambda}(x,x_2,\frac{3}{2}) \overline{\Lambda}(x,x_3,\frac{3}{2}+k) \overline{\Lambda}(x,x_4,\frac{3}{2}+k)  \\
&-\int \text{d}^3x \frac{\lambda^2}{4\pi a} \sum_{k=0}^{\infty}  D_k \frac{\text{d}}{\text{d}\epsilon}|_{\epsilon=0} (r_{34}^2)^{k+\epsilon} \overline{\Lambda}(x,x_1,\frac{3}{2}) \overline{\Lambda}(x,x_2,\frac{3}{2}) \overline{\Lambda}(x,x_3,\frac{3}{2}+k+\epsilon) \overline{\Lambda}(x,x_4,\frac{3}{2}+k+\epsilon) \\
=&\frac{\lambda^2}{4\pi a} \sum_{k=0}^{\infty} C_k (r_{34}^2)^k \mathcal{I}_4(\frac{3}{2},\frac{3}{2},\frac{3}{2}+k,\frac{3}{2}+k)  \\
&-\frac{\lambda^2}{4\pi a} \sum_{k=0}^{\infty}  D_k \frac{\text{d}}{\text{d}\epsilon}|_{\epsilon=0} (r_{34}^2)^{k+\epsilon} \mathcal{I}_4(\frac{3}{2},\frac{3}{2},\frac{3}{2}+k+\epsilon,\frac{3}{2}+k+\epsilon)\,.
\label{K4RecoverCoveriance}
  \end{aligned}
\end{equation}

As shown diagrammatically in (\ref{fig:schannel}), we managed to rewrite a four point, one-loop diagram as a series of tree level diagrams, reducing the loop number by 1 at the cost of now having an infinite series to sum over.
\begin{equation}
\raisebox{-0.7cm}{%
\begin{tikzpicture}
\draw[fill=black]
  (0,0) circle(2pt)
  -- +(120:0.6cm) circle(2pt) node[above left] {$\tfrac{3}{2}$}
  +(0,0) -- +(-120:0.6cm) circle(2pt) node[below left] {$\tfrac{3}{2}$}
  ++(0.4cm,0) circle(2pt)
  -- +(60:0.6cm) circle(2pt) node[above right] {$\tfrac{3}{2}$}
  +(0,0) -- +(-60:0.6cm) circle(2pt) node[below right] {$\tfrac{3}{2}$};

\draw (0.2cm,0cm) circle(0.75cm);
\draw (0cm,0cm) arc(180:0:0.2cm) (0.6cm,0);
\draw (0cm,0cm) arc(180:360:0.2cm) (0.6cm,0);
\end{tikzpicture}}
\raisebox{0.5cm}{$= \sum_{k\geq 0}C_k$}
\raisebox{-0.7cm}{\begin{tikzpicture}
\draw[fill=black]
			(0,0)circle(2pt) -- +(45:0.74cm) circle(2pt) node[above right] {$\tfrac{3}{2}+k$}
			+(0,0)  -- +(135:0.74cm) circle(2pt) node[above left] {$\tfrac{3}{2}$}
			+(0,0)  -- +(-135:0.74cm) circle(2pt) node[below left] {$\tfrac{3}{2}$}
			+(0,0) -- +(-45:0.74cm) circle(2pt) node[below right] {$\tfrac{3}{2}+k$};
			\draw (0cm,0cm) circle(0.75cm);
\end{tikzpicture}}
\raisebox{0.5cm}{$+\sum_{k\geq 0}D_k\frac{\text{d}}{\text{d}\epsilon}|_{\epsilon=0}$}
\raisebox{-0.7cm}{\begin{tikzpicture}
\draw[fill=black]
			(0,0)circle(2pt) -- +(45:0.74cm) circle(2pt) node[above right] {$\tfrac{3}{2}+k+\epsilon$}
			+(0,0)  -- +(135:0.74cm) circle(2pt) node[above left] {$\tfrac{3}{2}$}
			+(0,0)  -- +(-135:0.74cm) circle(2pt) node[below left] {$\tfrac{3}{2}$}
			+(0,0) -- +(-45:0.74cm) circle(2pt) node[below right] {$\tfrac{3}{2}+k+\epsilon$};
			\draw (0cm,0cm) circle(0.75cm);
\end{tikzpicture}}
\label{fig:schannel}
\end{equation}
This is the key equation which will allow us to extract contributions from the $t$- and $u$-channel bubble diagram to the conformal dimensions of the double trace operators in term terms of tree-level data. 

In order to demonstrate that the series expansion trick preserves conformal covariance, below we make use of the representation of $I_4$ in \cite{Dolan:2000uw},
\begin{equation}
  \begin{aligned}
I_4(\delta_1,\delta_2,\delta_3,\delta_4)=&\frac{1}{r_{13}^{2\mu}}\left(\frac{r_{14}}{r_{24}}\right)^{2\delta_2}\left(\frac{r_{14}}{r_{34}}\right)^{2\delta_3-2\mu}\left(\frac{r_{13}}{r_{34}}\right)^{2\delta_4} \\
&\times H(\delta_2,\mu-\delta_4,\delta_1+\delta_2+1-\mu,\delta_1+\delta_2;u,v)\\
=&\frac{1}{r_{13}^{2\mu}}\left(\frac{r_{13}}{r_{23}}\right)^{2\delta_2}\left(\frac{r_{12}}{r_{23}}\right)^{2\delta_3-2\mu}\left(\frac{r_{12}}{r_{24}}\right)^{2\delta_4} \\
&\times H(\delta_4,\mu-\delta_2,\delta_3+\delta_4+1-\mu,\delta_3+\delta_4;u,v),
\label{I4ExplicitExpression}
  \end{aligned}
\end{equation}
where $\mu=\frac{1}{2}(\delta_1+\delta_2+\delta_3+\delta_4)$ and $H$ is a conformal invariant quantity. For the $s$-channel we will use the first identity in (\ref{I4ExplicitExpression}) where the factor of $(r_{34}^2)^{k+\epsilon}$ in (\ref{K4RecoverCoveriance}) is precisely canceled by $r_{13}^{-3}r_{24}^{-3}r_{34}^{-2k-2\epsilon}$ in front of $H$, leaving us with $r_{12}^{-3} r_{34}^{-3} u^\frac{3}{2}$ which exhibits the correct covariant behavior of a conformal correlator. For the $t$- and $u$-channel the second identity is to be used and a similar argument follows. The explicit expression of $H$ is not immediately useful and we define them later in (\ref{Hfunction1stTerm}) and (\ref{Hfunction2ndTerm}).

\subsection{The $s$-channel as a cross check}
\label{SectionSchannelSpectralFunction}

From (\ref{K4RecoverCoveriance}) we can already extract dual CFT data of the $s$-channel diagram by reading off the spectral function. Of course, its  explicit expression for the $s$-channel has already  been determined in \cite{Carmi:2018qzm} through a bootstrap. In this section we intend to perform a crosscheck between the bootstrapped spectral function and our results obtained from perturbative calculation. A recap on spectral functions of AdS correlators and their relationship to the dual CFT is recapped in Appendix \ref{AppendixSpectralFunction}.

As in \cite{Sachs:2023eph} we rewrite the four point correlation function in terms of a spectral representation 
\begin{equation}\label{eq:WDg}
  \begin{aligned}
W(x_1,x_2,x_3,x_4)=\int_\mathbb{R} \text{d}\nu D(\nu) g^{\Delta,\Delta,\Delta,\Delta}_{x_1,x_2,x_3,x_4,\nu}\,,
  \end{aligned}
\end{equation}
where $D(\nu)$ is the 'spectral function'\footnote{As opposed to the actual Källén–Lehmann spectral representation of the two point function, $B(\nu)$, seen in (\ref{K-Lspecrep})} and $g^{\Delta,\Delta,\Delta,\Delta}_{x_1,x_2,x_3,x_4,\nu}$ is the conformal block with an intermediate double trace operator of dimension $\Delta(\nu)=\frac{d}{2}+i\nu$. The latter is related to the $s$-channel conformal partial wave
\begin{equation}
  \begin{aligned}
\Psi^{\Delta,\Delta,\Delta,\Delta}_{x_1,x_2,x_3,x_4,\nu} = \int_{\partial \text{AdS}_{d+1}} \text{d}^dx \langle \mathcal{O}_{\Delta}(x_1) \mathcal{O}_{\Delta}(x_2)\mathcal{O}_{\frac{d}{2}+i\nu}(x_0)\rangle \langle \mathcal{O}_{\frac{d}{2}-i\nu}(x_0) \mathcal{O}_{\Delta}(x_3) \mathcal{O}_{\Delta}(x_4) \rangle
\label{CPWdef}
  \end{aligned}
\end{equation}
by a normalizatiom, i.e.
\begin{equation}\label{eq:PsiDg}
  \begin{aligned}
\Psi^{\Delta,\Delta,\Delta,\Delta}_{x_1,x_2,x_3,x_4,\nu} = K(\nu) g^{\Delta,\Delta,\Delta,\Delta}_{x_1,x_2,x_3,x_4,\nu},
  \end{aligned}
\end{equation}
\begin{equation}\label{eq:K}
  \begin{aligned}
K(\nu)=\frac{\pi^{\frac{d}{2}}\Gamma(-i\nu)\Gamma(\frac{d}{4}+i\frac{\nu}{2})^2}{\Gamma(\frac{d}{2}+i\nu)\Gamma(\frac{d}{4}-i\frac{\nu}{2})}.
  \end{aligned}
\end{equation}
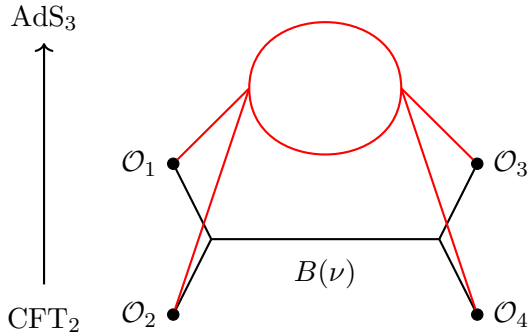
\begin{figure}[t]
\begin{tikzpicture}
  \begin{feynman}
    \vertex (a) at (-1.5,0);
    \vertex (b) at (1.5,0);
    \node [dot,label=left:\(\mathcal{O}_1\)] (o1) at (-2,1);
    \node [dot,label=left:\(\mathcal{O}_2\)] (o2) at (-2,-1);
    \node [dot,label=right:\(\mathcal{O}_3\)] (o3) at (2,1);
    \node [dot,label=right:\(\mathcal{O}_4\)] (o4) at (2,-1);
    \node [label=\(\text{CFT}_2\)] (CFT2) at (-3.7,-1.5);
    \node [label=\(\text{AdS}_3\)] (AdS3) at (-3.7,2.5);
    \node [label=below:\(B(\nu)\)] (doubletrace) at (0,0);
    \vertex (c) at (-1,2);
    \vertex (d) at (1,2);
    \vertex (arrowa) at (-3.7,-1);
    \vertex (arrowb) at (-3.7,3);

    \diagram* {
      (a) -- [thick] (b),
      (o1) -- [thick] (a),
      (o2) -- [thick] (a),
      (o3) -- [thick] (b),
      (o4) -- [thick] (b),
      (o1) -- [thick, red] (c),
      (o2) -- [thick, red] (c),
      (o3) -- [thick, red] (d),
      (o4) -- [thick, red] (d),
      (c) -- [half left, thick, red] (d) -- [half left, thick, red] (c),
      
    };

    \draw[->, thick] ($ (arrowa)!0.1!(arrowb) $) to ($ (arrowa)!0.9!(arrowb) $) ;
  \end{feynman}
\end{tikzpicture}
\centering
\label{DoubleTraceSpecRep}
\caption{The intermediate double trace operator dual to a bubble diagram in the bulk}
\end{figure}
In (\ref{CPWdef}), explicit expression of the CFT 3-point function reads
\begin{equation}
  \begin{aligned}
\langle \mathcal{O}_{\Delta}(x_i) \mathcal{O}_{\Delta}(x_i)\mathcal{O}_{\frac{d}{2}+i\nu}(x_0)\rangle = \frac{1}{x_{ij}^{2\Delta-\Delta(\nu)}x_{i0}^{\Delta(\nu)}x_{j0}^{\Delta(\nu)}}.
  \end{aligned}
\end{equation}
which is related to the field theory correlator by a normalization $b_{\Delta,\Delta,\nu}$ defined in (\ref{threepointfunction}). Moreover, a reconstruction of the cross diagram in terms of the conformal partial wave is given in (\ref{CrossDiagramAsCPW}). Therefore, reconstructing each individual cross diagram appearing in the sum  (\ref{K4RecoverCoveriance}) yields
\begin{equation}
  \begin{aligned}
\mathcal{K}_4^s=& \frac{\lambda^2}{4\pi a} \sum_{k=0}^{\infty} C_k (r_{34}^2)^k \int_\mathbb{R} \text{d}\nu \frac{\nu^2}{\pi} b_{\Delta,\Delta,\nu} b_{\Delta+k,\Delta+k,-\nu} \Psi^{\Delta,\Delta,\Delta+k,\Delta+k}_{x_1,x_2,x_3,x_4,\nu}  \\
&-\frac{\lambda^2}{4\pi a} \sum_{k=0}^{\infty}  D_k \frac{\text{d}}{\text{d}\epsilon}|_{\epsilon=0} (r_{34}^2)^{k+\epsilon} \int_\mathbb{R} \text{d}\nu \frac{\nu^2}{\pi} b_{\Delta,\Delta,\nu} b_{\Delta+k+\epsilon,\Delta+k+\epsilon,-\nu} \Psi^{\Delta,\Delta,\Delta+k+\epsilon,\Delta+k+\epsilon}_{x_1,x_2,x_3,x_4,\nu} \\
=& \frac{\lambda^2}{4\pi a} \sum_{k=0}^{\infty} C_k \int_\mathbb{R} \text{d}\nu \frac{\nu^2}{\pi} b_{\Delta,\Delta,\nu} b_{\Delta+k,\Delta+k,-\nu} \Psi^{\Delta,\Delta,\Delta,\Delta}_{x_1,x_2,x_3,x_4,\nu}  \\
&-\frac{\lambda^2}{4\pi a} \sum_{k=0}^{\infty}  D_k \frac{\text{d}}{\text{d}\epsilon}|_{\epsilon=0} \int_\mathbb{R} \text{d}\nu \frac{\nu^2}{\pi} b_{\Delta,\Delta,\nu} b_{\Delta+k+\epsilon,\Delta+k+\epsilon,-\nu} \Psi^{\Delta,\Delta,\Delta,\Delta}_{x_1,x_2,x_3,x_4,\nu},
  \end{aligned}
\end{equation}
where in the last step we used the following relation between conformal partial waves:
\begin{equation}
  \begin{aligned} &\Psi^{\Delta,\Delta,\Delta+k+\epsilon,\Delta+k+\epsilon}_{x_1,x_2,x_3,x_4,\nu} \\
  =&\int_{\partial \text{AdS}_{d+1}} \text{d}^dx_0 \langle \mathcal{O}_{\Delta}(x_1) \mathcal{O}_{\Delta}(x_2)\mathcal{O}_{\frac{d}{2}+i\nu}(x_0)\rangle \langle \mathcal{O}_{\frac{d}{2}-i\nu}(x_0) \mathcal{O}_{\Delta+k+\epsilon}(x_3) \mathcal{O}_{\Delta+k+\epsilon}(x_4) \rangle \\
  =&\int_{\partial \text{AdS}_{d+1}} \text{d}^dx_0 \frac{1}{x_{12}^{2\Delta-\Delta(\nu)}x_{10}^{\Delta(\nu)}x_{20}^{\Delta(\nu)}} \frac{1}{x_{34}^{2\Delta+2k+2\epsilon-\Delta(\nu)}x_{30}^{\Delta(\nu)}x_{40}^{\Delta(\nu)}} \\
  =&\frac{1}{(r_{34}^2)^{k+\epsilon}}\Psi^{\Delta,\Delta,\Delta,\Delta}_{x_1,x_2,x_3,x_4,\nu}.
  \label{CPWbasis}
  \end{aligned}
\end{equation}
It is then easy to read off
\begin{equation}
  \begin{aligned}
B^{\text{perturb}}(\nu)=& \frac{1}{2} \frac{\lambda}{4\pi a} \sum_{k=0}^{\infty} \left( C_k (r_{34}^2)^k \frac{b_{\Delta+k,\Delta+k,-\nu}}{b_{\Delta,\Delta,\nu}} -D_k \frac{\text{d}}{\text{d}\epsilon}|_{\epsilon=0} (r_{34}^2)^{k+\epsilon} \frac{b_{\Delta+k+\epsilon,\Delta+k+\epsilon,-\nu}}{b_{\Delta,\Delta,\nu}}\right) \\
\defeq& \sum_{k=0}^{\infty} B^k(\nu),
\label{spectralfunctionfromellipticexpansion}
  \end{aligned}
\end{equation}
where the $\frac{1}{2}$ came from the symmetry factor of the one-loop diagram. 

On the other hand, through bootstrap it was found in \cite{Carmi:2018qzm} that for $d=2$,
\begin{equation}
  \begin{aligned}
B^{\text{bootstrap}}(\nu)=& \frac{\lambda}{2}\frac{i\left( \Psi(\Delta-\frac{1+i\nu}{2})-\Psi(\Delta-\frac{1-i\nu}{2})\right)}{8\pi\nu}
\label{spectralfunctionbootstrapped}
  \end{aligned}
\end{equation}
in which $\Psi$ is the digamma function. Our next goal is to show that (\ref{spectralfunctionfromellipticexpansion}) and (\ref{spectralfunctionbootstrapped}) are equivalent and give the same CFT data on the boundary.

In Appendix \ref{AppendixSpectralFunction}, we derived the following expression of anomalous dimension and OPE coefficient depending on the spectral functions,
\begin{equation}
  \begin{aligned}
\gamma^{(2)}_{n,0} =&\gamma^{(1)}_{n,0}B^{(0)}_{\nu_n},
  \end{aligned}
\end{equation}
and
\begin{equation}
  \begin{aligned}
A^{(2)}_{n,0} =& \frac{2\pi i}{(\gamma^{(1)}_{n,0})^2}(D^{\times^{(0)}}_{\nu_n} B^{(-1)}_{\nu_n} + D^{\times^{(-2)}}_{\nu_n}B^{(1)})_{\nu_n}.
  \end{aligned}
\end{equation}
Therefore, it is sufficient to show that the two spectral functions have the same Laurent series coefficients in the -1, 0 and 1st order around $\nu_n=\Delta+n-\frac{d}{4}$. Indeed for a fixed $\nu_n$,
\begin{equation}
  \begin{aligned}
{B^k}^{(-1,0)}_{\nu_n}=0, \text{ if } k>n.
  \end{aligned}
\end{equation}
Hence by summing up only a finite number of $k$, it is possible to check explicitly that
\begin{equation}
  \begin{aligned}
&{B^{\text{perturb}}}^{(-1,0)}(\nu_n)=\sum_{k=0}^{n}{B^k}^{(-1,0)}_{\nu_n}={B^{\text{bootstrap}}}^{(-1,0)}(\nu_n).
\label{B-10consistencycheck}
  \end{aligned}
\end{equation}
Since $\gamma^{(2)}_{n,0}$ depends only on $B^{(0)}_{\nu_n}$, one can thus extract the anomalous dimension from a finite number of terms, even though $B^{\text{perturb}}(\nu)$ takes the form of an infinite sum. The $s$-channel anomalous dimensions is then found to be
\begin{equation}
  \begin{aligned}
{\gamma^{(2)}}^s_{n,0}=-\frac{\lambda^2}{256\pi^2(n+1)^3}.
\label{AnomDimS}
  \end{aligned}
\end{equation}

Concerning the OPE coefficients, we note that ${B^k}^{(1)}_{\nu_n}$ are nonzero at all orders of $k$ and $n$, so that one does have to solve the infinite sum in order to extract $A^{(2)}_{n,0}$ from ${B^{\text{perturb}}}^{(1)}(\nu_n)=\sum_{k=0}^{\infty}{B^k}^{(1)}_{\nu_n}$. Nevertheless, by summing up the first few hundreds of terms numerically, we found the approximation to be in agreement with analytic results given by ${B^{\text{bootstrap}}}^{(1)}(\nu_n)$. The first few entries of $A^{(2)}_{n,0}$ are shown in Table \ref{OPECoeffS}.

\begin{table}[h!]
\centering
\setlength{\extrarowheight}{2pt}
\begin{tabular}{
| >{\centering\arraybackslash}m{1cm}
| >{\centering\arraybackslash}m{10cm} |
}
\hline
 & ${A^{(2)}_{n,0}}^s$ \\
\hline
n=0 &
$-\frac{11}{2}-\frac{\pi^2}{6}
+20\log 2
-16\log^2 2$ \\
\hline
n=1 &
$-\frac{163}{128}-\frac{3\pi^2}{32}
+\frac{69}{8}\log 2
-9\log^2 2$ \\
\hline
n=2 &
$-\frac{23751}{131072}-\frac{675\pi^2}{32768}
+\frac{1755}{1024}\log 2
-\frac{2025}{1024}\log^2 2$ \\
\hline
n=3 &
$-\frac{394619}{18874368}-\frac{1225\pi^2}{393216}
+\frac{48335}{196608}\log 2
-\frac{1225}{4096}\log^2 2$ \\
\hline
n=4 &
$-\frac{73522975}{34359738368}
-\frac{826875\pi^2}{2147483648}
+\frac{15789375}{536870912}\log 2
-\frac{2480625}{67108864}\log^2 2$ \\
\hline
\end{tabular}
\caption{First few OPE coefficients at 2nd order}
\label{OPECoeffS}
\end{table}

At this point all are settled with the $s$-channel. By expanding the elliptic integral we were able to reduce the loop diagram into a sum of tree level diagrams and hence read off its spectral function, due to the close relation between conformal partial wave and the tree level cross diagram. We managed to reproduce existing results of the boundary CFT, although our method, having to deal with a finite (in the case of anomalous dimensions) and infinite (in the case of OPE coefficients) sum, is in no way better than \cite{Carmi:2018qzm}. However, as we shall see soon, in the $t$- and $u$-channel there is no easy way to read off the spectral function. That is when the expansion trick stands out and allow us to obtain similar results through conformal block expansion.

\section{Conformal block expansion in the $t$- and $u$-channel}\label{sec:t-u}
Unfortunately, the methodology employed for the $s$-channel is not directly applicable to the $t$- and $u$-channels. In \cite{Carmi:2018qzm}, the spectral function in the $s$-channel was bootstrapped using constraints arising from disconnected diagrams. These constraints do not encode information about double-trace operators with spin and are therefore insufficient for bootstrapping the $t$- and $u$-channel spectral functions, which involve nontrivial spin exchange. In principle, one may obtain the $t$- and $u$-channel CFT data from the $s$-channel data via the use of 
6j symbol (e.g.\cite{Meltzer:2019nbs}); however, the resulting expression proved highly tedious that it was unclear to us how explicit results can be extracted. We begin this section by demonstrating that, even when employing the expansion trick (\ref{K4RecoverCoveriance}), the spectral function cannot be read off directly. We then proceed in Section \ref{SectionCBExapansionIntro} with an introduction to an algorithm of conformal block expansion, which so far appears to be the only method that allows for a successful extraction of the dual CFT data in the $t$- and $u$-channels. In Section \ref{SectionSchannelCBExpansion} conformal block expansion is performed for the $s$-channel as a consistency check against spectral function. Section \ref{SectionTUchannelCBExpansion} follows the same procedure to derive the $t$- and $u$-channel contributions to the anomalous dimensions.
\subsection{Spectral function does not work easily for $t$- and $u$-channel}

Much like (\ref{K4RecoverCoveriance}), the $t$-channel loop diagram reads

\begin{equation}
  \begin{aligned}
\mathcal{K}_4^{t}=&\frac{\lambda^2}{4\pi a} \sum_{k=0}^{\infty} C_k (r_{24}^2)^k \mathcal{I}_4(\frac{3}{2},\frac{3}{2}+k,\frac{3}{2},\frac{3}{2}+k)  \\
&-\frac{\lambda^2}{4\pi a} \sum_{k=0}^{\infty}  D_k \frac{\text{d}}{\text{d}\epsilon}|_{\epsilon=0} (r_{24}^2)^{k+\epsilon} \mathcal{I}_4(\frac{3}{2},\frac{3}{2}+k+\epsilon,\frac{3}{2},\frac{3}{2}+k+\epsilon) \\
=& \frac{\lambda^2}{4\pi a} \sum_{k=0}^{\infty} C_k (r_{24}^2)^k \int_\mathbb{R} \text{d}\nu \frac{\nu^2}{\pi} b_{\Delta,\Delta+k,\nu} b_{\Delta+k,\Delta,-\nu} \Psi^{\Delta,\Delta+k,\Delta,\Delta+k}_{x_1,x_2,x_3,x_4,\nu}  \\
&-\frac{\lambda^2}{4\pi a} \sum_{k=0}^{\infty}  D_k \frac{\text{d}}{\text{d}\epsilon}|_{\epsilon=0} (r_{24}^2)^{k+\epsilon} \int_\mathbb{R} \text{d}\nu \frac{\nu^2}{\pi} b_{\Delta,\Delta+k+\epsilon,\nu} b_{\Delta+k+\epsilon,\Delta,-\nu} \Psi^{\Delta,\Delta+k+\epsilon,\Delta,\Delta+k+\epsilon}_{x_1,x_2,x_3,x_4,\nu},
  \end{aligned}
\end{equation}

or diagrammatically

\begin{equation}
\raisebox{-0.7cm}{\begin{tikzpicture}[rotate=90]
\draw[fill=black](0,0)circle(2pt) 
-- +(120:0.6cm) circle(2pt) node[below left] {$\tfrac{3}{2}$}
 +(0,0)  --  +(-120:0.6cm) circle(2pt) node[below right] {$\tfrac{3}{2}$} 
 ++(0.4cm,0) circle(2pt) --  +(60:0.6cm) circle(2pt) node[above left] {$\tfrac{3}{2}$}
 +(0,0) -- +(-60:0.6cm) circle(2pt) node[above right] {$\tfrac{3}{2}$};
 	\draw (0.2cm,0cm) circle(0.75cm);
\draw (0cm,0cm) arc(180:0:0.2cm)  (0.6cm,0);
\draw (0cm,0cm) arc(180:360:0.2cm)  (0.6cm,0);
\end{tikzpicture}}
\raisebox{0.5cm}{$= \sum_{k\geq 0}C_k$}
\raisebox{-0.7cm}{\begin{tikzpicture}
\draw[fill=black]
			(0,0)circle(2pt) -- +(45:0.74cm) circle(2pt) node[above right] {$\tfrac{3}{2}$}
			+(0,0)  -- +(135:0.74cm) circle(2pt) node[above left] {$\tfrac{3}{2}$}
			+(0,0)  -- +(-135:0.74cm) circle(2pt) node[below left] {$\tfrac{3}{2}+k$}
			+(0,0) -- +(-45:0.74cm) circle(2pt) node[below right] {$\tfrac{3}{2}+k$};
			\draw (0cm,0cm) circle(0.75cm);
\end{tikzpicture}}
\raisebox{0.5cm}{$+\sum_{k\geq 0}D_k\frac{\text{d}}{\text{d}\epsilon}|_{\epsilon=0}$}
\raisebox{-0.7cm}{\begin{tikzpicture}
\draw[fill=black]
			(0,0)circle(2pt) -- +(45:0.74cm) circle(2pt) node[above right] {$\tfrac{3}{2}$}
			+(0,0)  -- +(135:0.74cm) circle(2pt) node[above left] {$\tfrac{3}{2}$}
			+(0,0)  -- +(-135:0.74cm) circle(2pt) node[below left] {$\tfrac{3}{2}+k+\epsilon$}
			+(0,0) -- +(-45:0.74cm) circle(2pt) node[below right] {$\tfrac{3}{2}+k+\epsilon$};
			\draw (0cm,0cm) circle(0.75cm);
\end{tikzpicture}}
\label{fig:tchannel}.
\end{equation}

However, a problem arises that

\begin{equation}
  \begin{aligned} &\Psi^{\Delta,\Delta+k+\epsilon,\Delta,\Delta+k+\epsilon}_{x_1,x_2,x_3,x_4,\nu} \\
  =&\int_{\partial \text{AdS}_{d+1}} \text{d}^dx_0 \langle \mathcal{O}_{\Delta}(x_1) \mathcal{O}_{\Delta+k+\epsilon}(x_2)\mathcal{O}_{\frac{d}{2}+i\nu}(x_0)\rangle \langle \mathcal{O}_{\frac{d}{2}-i\nu}(x_0) \mathcal{O}_{\Delta}(x_3) \mathcal{O}_{\Delta+k+\epsilon}(x_4) \rangle \\
  =&\int_{\partial \text{AdS}_{d+1}} \text{d}^dx_0 \frac{1}{x_{12}^{2\Delta+k+\epsilon-\Delta(\nu)}x_{10}^{\Delta(\nu)}x_{20}^{\Delta(\nu)}} \frac{1}{x_{34}^{2\Delta+k+\epsilon-\Delta(\nu)}x_{30}^{\Delta(\nu)}x_{40}^{\Delta(\nu)}} \\
  \neq &\frac{1}{(r_{24}^2)^{k+\epsilon}}\Psi^{\Delta,\Delta,\Delta,\Delta}_{x_1,x_2,x_3,x_4,\nu}.
  \end{aligned}
\end{equation}
Thus, if we consider $\Psi^{\Delta,\Delta,\Delta,\Delta}_{x_1,x_2,x_3,x_4,\nu}$ and $\Psi^{\Delta,\Delta+k+\epsilon,\Delta,\Delta+k+\epsilon}_{x_1,x_2,x_3,x_4,\nu}$ each as an infinite dimensional set of basis indexed by $\nu$ spanning the space of all four point correlators, there is no diagonal transformation matrix between these two sets of basis, unlike the case of $\Psi^{\Delta,\Delta,\Delta+k+\epsilon,\Delta+k+\epsilon}$ in the $s$-channel. Hence we cannot easily read off the spectral function while having the $s$-channel conformal partial waves as basis, which is a prerequisite for obtaining dual CFT information with respect to the $s$-channel conformal block. Alternatively, $\Psi^{\Delta,\Delta+k+\epsilon,\Delta,\Delta+k+\epsilon}_{x_1,x_2,x_3,x_4,\nu}$ is indeed related to the $t$-channel conformal partial waves, \begin{equation}
  \begin{aligned}
\Psi^{t\text{-channel},\Delta,\Delta,\Delta,\Delta}_{x_1,x_2,x_3,x_4,\nu} = \int_{\partial \text{AdS}_{d+1}} \text{d}^dx \langle \mathcal{O}_{\Delta}(x_1) \mathcal{O}_{\Delta}(x_3)\mathcal{O}_{\frac{d}{2}+i\nu}(x_0)\rangle \langle \mathcal{O}_{\frac{d}{2}-i\nu}(x_0) \mathcal{O}_{\Delta}(x_2) \mathcal{O}_{\Delta}(x_4) \rangle
\label{CPWdeftchannel}
  \end{aligned}
\end{equation}
via a diagonal matrix, taking the limit $x_1\rightarrow x_3$ and $x_2\rightarrow x_4$. It is then possible to convert between $t$- and $s$-channel conformal partial waves using the conformal 6j symbol \cite{Meltzer:2019nbs} \cite{Liu:2018jhs}, following a similar approach as in \cite{Dujava:2025php}[3.23]. However, we found it challenging to extract explicit results of the CFT data, as the subsequent expressions contain products of generalized hypergeometric functions which proved hard to evaluate analytically. For this reason we resort to the method of conformal block expansion as performed in \cite{Bertan:2018afl,Heckelbacher:2020nue,Heckelbacher:2022fbx,Heckelbacher:2022fzi}.

\subsection{Conformal Block Expansion}

\label{SectionCBExapansionIntro}
Since, unlike of the $s$-channel, the spectral function for the $t$- and $u$-channel cannot be read off directly form the respective expressions of the bubble Feynman integrals, in this section we will resort to the method of conformal block expansion as performed in \cite{Bertan:2018afl} \cite{Heckelbacher:2020nue} \cite{Heckelbacher:2022fbx} \cite{Heckelbacher:2022fzi}.
Comparison between (\ref{eq:CBExp}) and 
\begin{align}
    \begin{split}
\label{4ptftofirstorder}
    \langle \bar{\phi}(x_1)& \bar{\phi}(x_2) \bar{\phi}(x_3) \bar{\phi}(x_4) \rangle = \diagramtwob  \\[0.5em]
    &+\lambda~ \diagramtwoc + \mathcal{O}\left( \lambda^3 \right)\\[0.5em]
    &\defeq \frac{N_\phi^2}{(r_{12}r_{34})^{2\Delta}} \left[1+  u^\Delta  + \frac{u^\Delta}{v^\Delta} +\lambda \mathcal{I}_4 +\frac{\lambda^2}{2} \left( \mathcal{K}_4^s +\mathcal{K}_4^t +\mathcal{K}_4^u \right) \right]+ \mathcal{O}\left( \lambda^3 \right)\mkern-3mu,
        \end{split}
\end{align}
enables us to extract $\gamma^{(1)}_{n,l}$, $\gamma^{(2)}_{n,l}$, $A^{(1)}_{n,l}$ and $A^{(2)}_{n,l}$. Individual terms are compared in the following order:
\begin{enumerate}
    \item $W^{\times}_{\mathcal{O}(\log u)}\defeq \left( \gamma^{(1)}_{n,l}A_{n,l}G'_{n,l} \right)_{\mathcal{O}(\log u)}$, starting from $n=l=0$. $\gamma^{(1)}_{n,l}$ could then be determined recursively. To determine a higher order $\gamma^{(1)}_{N,L}$, knowledge from previous iterations of all $\gamma^{(1)}_{n,l}$ with $n<N$ and $l<L$ is required. More details on this recursive algorithm can be found in Appendix \ref{AppendixCBDef}.

    \item $A^{(1)}_{n,l}$ are determined recursively by matching $W^{\times}_{\mathcal{O}(1)}$ to the remaining terms in the conformal block expansion that is independent of $\log u$,
    \begin{equation*}
        W^{\times}_{\mathcal{O}(1)}=\left( \gamma^{(1)}_{n,l}(\Delta)\left(A_{n,l}G'_{n,l}+A^{(1)}_{n,l}G_{n,l}\right) \right)_{\mathcal{O}(1)}.
    \end{equation*}

    \item $W^{\rangle\!\circ\!\langle}_{\mathcal{O}(\log^2 u)}\defeq \left( \frac{1}{2}(\gamma^{(1)}_{n,l})^2A_{n,l}G''_{n,l} \right)_{\mathcal{O}(\log^2 u)}$, so that $\gamma^{(1)}_{n,l}$ is again determined independently. This provides us with a consistency check, which is equivalent to (\ref{consistencycondition}) and \cite[4.24]{Carmi:2018qzm}, that enables the spectral function of the loop diagram to be bootstrapped.

    \item $W^{\rangle\!\circ\!\langle}_{\mathcal{O}(\log u)}\defeq \left( \gamma^{(2)}_{n,l}A_{n,l}G'_{n,l} \right)_{\mathcal{O}(\log u)}$, so that $\gamma^{(2)}_{n,l}$ are determined recursively.

    \item $A^{(2)}_{n,l}$ are determined recursively by matching $W^{\rangle\!\circ\!\langle}_{\mathcal{O}(1)}$ to the remaining terms in the conformal block expansion that is independent of $\log u$,
    \begin{equation*}
    \begin{aligned}
        W^{\rangle\!\circ\!\langle}_{\mathcal{O}(1)}=&\Bigg( \frac{1}{2}(\gamma^{(1)}_{n,l}(\Delta))^2\left(A_{n,l}G''_{n,l}+A^{(2)}_{n,l}G_{n,l}+2A^{(1)}_{n,l}G'_{n,l}\right)\cr
	    &+\gamma^{(2)}_{n,l}(\Delta)\left(A_{n,l}G'_{n,l}+A^{(1)}_{n,l}G_{n,l}\right)\Bigg)_{\mathcal{O}(1)}.
    \end{aligned}
    \end{equation*}

\end{enumerate}
Furthermore, since $\gamma^{(2)}_{n,l}$ and $A^{(2)}_{n,l}$ appear in the conformal block expansion only up to linear order, it is possible to split them into

\begin{equation}
\begin{aligned}
    \gamma^{(2)}_{n,l}=\gamma^{(2)s}_{n,l}+\gamma^{(2)t}_{n,l}+\gamma^{(2)u}_{n,l}
\end{aligned}
\end{equation}
and
\begin{equation}
\begin{aligned}
    A^{(2)}_{n,l}=A^{(2)s}_{n,l}+A^{(2)t}_{n,l}+A^{(2)u}_{n,l},
\end{aligned}
\label{AnomDimSpliting}
\end{equation}
so that loop corrections in $s$-, $t$- and $u$-channels correspond with the conformal block expansion,

\begin{equation}
\begin{aligned}
    W^{\rangle\!\circ\!\langle}_{s}\defeq&\frac{1}{2}(\gamma^{(1)}_{n,l})^2\left(A_{n,l}G''_{\Delta(n,l),l}+A^{(2)s}_{n,l}G_{\Delta(n,l),l}+2A^{(1)}_{\Delta(n,l),l}G'_{\Delta(n,l),l}\right)\cr
	&+(\gamma^{(2)s}_{n,l})\left(A_{n,l}G'_{\Delta(n,l),l}+A^{(1)}_{n,l}G_{\Delta(n,l),l}\right)
    \label{gamma2s}
\end{aligned}
\end{equation}
and
\begin{equation}
\begin{aligned}
    W^{\rangle\!\circ\!\langle}_{t,u} \defeq &\frac{1}{2}(\gamma^{(1)}_{n,l})^2 A^{(2)t,u}_{n,l}G_{\Delta(n,l),l}\cr
	&+(\gamma^{(2)t,u}_{n,l})\left(A_{n,l}G'_{\Delta(n,l),l}+A^{(1)}_{n,l}G_{\Delta(n,l),l}\right).
     \label{gamma2tu}
\end{aligned}
\end{equation}
The spiting of $\gamma^{(2)}_{n,l}$ and $A^{(2)}_{n,l}$ creates redundant degrees of freedom. As seen in (\ref{gamma2s}) and (\ref{gamma2tu}), we decided to have $A^{(2)s}$ and $\gamma^{(2)s}_{n,l}$ compensate for $\frac{1}{2}(\gamma^{(1)}_{n,l})^2 \big( A_{n,l}G''_{\Delta(n,l),l} + 2A^{(1)}_{\Delta(n,l),l}G'_{\Delta(n,l),l} \big)$ entirely, so that they fit also with \cite{Carmi:2018qzm} for an $\mathcal{O}(\mathcal{N})$ model at large $\mathcal{N}$ limit. However, one is free to redistribute this compensation among all 3 channels.

\subsection{The $s$-channel as a warm-up}

\label{SectionSchannelCBExpansion}

As a first example of conformal block expansion we begin again with the $s$-channel diagram

Since $\Delta_1+\Delta_2\neq\Delta_3+\Delta_4$, we insert (\ref{cross}) and \cite[B.11]{Dolan:2000uw} into (\ref{K4RecoverCoveriance}) and find 
\begin{equation}
  \begin{aligned}
\mathcal{K}_4^s=&\frac{\lambda^2}{4\pi a} \sum_{k=0}^{\infty} C_k (r_{34}^2)^k \mathcal{I}_4(\frac{3}{2},\frac{3}{2},\frac{3}{2}+k,\frac{3}{2}+k)  \\
&-\frac{\lambda^2}{4\pi a} \sum_{k=0}^{\infty}  D_k \frac{\text{d}}{\text{d}\epsilon}|_{\epsilon=0} (r_{34}^2)^{k+\epsilon} \mathcal{I}_4(\frac{3}{2},\frac{3}{2},\frac{3}{2}+k+\epsilon,\frac{3}{2}+k+\epsilon) \\
=&\lim_{\epsilon \rightarrow 0} \bigg( \frac{\lambda^2}{4\pi a} \sum_{k=0}^{\infty} C_k \frac{1}{(r_{13})^3(r_{24})^3}\frac{\pi}{\Gamma(\frac{3}{2})^2\Gamma(\frac{3}{2}+k)^2}\left(\frac{a}{2\pi}\right)^4 \frac{\Gamma(3+k-1)}{2}(H_1^{k+\epsilon}+H_2^{k+\epsilon})  \\
&-\frac{\lambda^2}{4\pi a} \sum_{k=0}^{\infty}  D_k \frac{\text{d}}{\text{d}\epsilon} \frac{1}{(r_{13})^3(r_{24})^3}\frac{\pi}{\Gamma(\frac{3}{2})^2\Gamma(\frac{3}{2}+k+\epsilon)^2}\left(\frac{a}{2\pi}\right)^4 \frac{\Gamma(3+k+\epsilon-1)}{2}(H_1^{k+\epsilon}+H_2^{k+\epsilon}) \bigg)\,,
  \end{aligned}
\end{equation}
where we recall from \eqref{I4ExplicitExpression} and used its first equality, with $H(\alpha,\beta,\gamma,\delta;u,v)=H_1(\alpha,\beta,\gamma,\delta;u,v)+H_2(\alpha,\beta,\gamma,\delta;u,v)$ whose explicit expression is given by \cite[5.9]{Dolan:2000uw}. Concretely we have
\begin{equation}
  \begin{aligned}
H_1^{k+\epsilon}&=H_1\!\left(\tfrac32,\tfrac32,1-k-\epsilon,3;u,v\right)=
\frac{\Gamma(k+\epsilon)}{2}
\bigl(\Gamma\!\left(\tfrac32\right)\bigr)^4\,
G\!\left(\tfrac32,\tfrac32,1-k-\epsilon,3;u,1-v\right), \\[6pt]
&\defeq \sum_{m,n=0}^\infty H_1^{k+\epsilon,m,n} u^m (1-v)^n
  \end{aligned}
  \label{Hfunction1stTerm}
\end{equation}
and
\begin{equation}
  \begin{aligned}
H_2^{k+\epsilon}&=H_2\!\left(\tfrac32,\tfrac32,1-k-\epsilon,3;u,v\right) \\
&=\frac{\Gamma(-k-\epsilon)}{\Gamma(3+2k+2\epsilon)}
\bigl(\Gamma\!\left(\tfrac32+k+\epsilon\right)\bigr)^2
\bigl(\Gamma\!\left(\tfrac72+k+\epsilon\right)\bigr)^2\\
&\qquad\qquad\times
u^{k+\epsilon}\,
G\!\left(\tfrac32+k+\epsilon,\tfrac32+k+\epsilon,
1+k+\epsilon,3+2k+2\epsilon;u,1-v\right) \\
&\defeq \sum_{m,n=0}^\infty u^{k+\epsilon} \times H_2^{k+\epsilon,m,n} u^m (1-v)^n\,,
  \end{aligned}
  \label{Hfunction2ndTerm}
\end{equation}
while the function $G$ is defined in \cite[5.1]{Dolan:2000uw} as
\begin{equation}
  \begin{aligned}
G(\alpha,\beta,\gamma,\delta;x,y) = \sum_{m,n=0} {(\delta-\alpha)_m
(\delta-\beta)_m \over m! (\gamma)_m} \, 
{(\alpha)_{m+n} (\beta)_{m+n} \over n! (\delta)_{2m+n}} \, x^m y^n.
  \end{aligned}
\end{equation}
The factor of $u^{k+\epsilon}$ in (\ref{Hfunction2ndTerm}) is of particular importance here. As this factor increases the power of $u$ by $k$ after taking the $\epsilon \rightarrow 0$ limit, for any fixed $i$ and $j$, contributions from $H_2^{k+\epsilon}$ to the $u^i(1-v)^j$ term in $\mathcal{K}_4^s$ could only come from those $H_2^{k+\epsilon,m,n}$ with $m+k<i$. Hence we rewrite
\begin{equation}
  \begin{aligned}
\mathcal{K}_4^s=&\lim_{\epsilon\rightarrow 0}\frac{\lambda^2}{128\pi^4} \frac{1}{(r_{13})^3(r_{24})^3}\sum_{i,j=0}^\infty \Bigg(H_{\log}^{i,j} + H_{\text{nonlog}}^{i,j} \Bigg) u^i (1-v)^j
  \end{aligned}
\end{equation}
where
\begin{equation}
  \begin{aligned}
H_{\log}^{i,j}=\sum_{k=0}^i \bigg( &C_k \frac{\Gamma(2+k)}{\Gamma(\frac{3}{2}+k+\epsilon)^2} (H_1^{k+\epsilon,i,j}+H_2^{k+\epsilon,i-k,j}) \\
+&D_k \frac{\text{d}}{\text{d}\epsilon} \frac{\Gamma(2+k+\epsilon)}{\Gamma(\frac{3}{2}+k+\epsilon)^2} (H_1^{k+\epsilon,i,j}+H_2^{k+\epsilon,i-k,j}) \bigg),
  \end{aligned}
\end{equation}
and
\begin{equation}
  \begin{aligned}
H_{\text{nonlog}}^{i,j}=\sum_{k=i+1}^\infty C_k \frac{\Gamma(2+k)}{\Gamma(\frac{3}{2}+k+\epsilon)^2} H_1^{k+\epsilon,i,j}.
  \end{aligned}
\end{equation}
As the $\epsilon\rightarrow 0$ limit is taken, the finite sum $H_{\log}^{i,j}$ gives rise to terms proportional to $\log u$ and $\log^2 u$, while the infinite sum $H_{\text{nonlog}}^{i,j}$ contributes to terms without $\log u$. The anomalous dimension is determined by the $\log u$ and $\log^2 u$ terms only, while the OPE coefficient is determined by terms without $\log u$ too. This is consistent with the observation in (\ref{B-10consistencycheck}) that the anomalous dimensions can be written as a finite sum, but the OPE coefficients cannot. Plugging the explicit expression of $\mathcal{K}_4^s$ back into (\ref{eq:CBExp}), we obtained again the exact values of the anomalous dimensions (\ref{AnomDimS}), and the  OPE coefficients by numeric approximation  in Table \ref{OPECoeffS}. 

\subsection{$t$- and $u$-channel anomalous dimensions}

\label{SectionTUchannelCBExpansion}

For the $t$- and $u$-channel, $\Delta_1+\Delta_2=\Delta_3+\Delta_4$, or in the convention of \cite[5.13]{Dolan:2000uw}, $\gamma=1$. We therefore plug in instead \cite[5.13]{Dolan:2000uw} into (\ref{K4RecoverCoveriance}), and use the second identity in (\ref{I4ExplicitExpression}), resulting in
\begin{equation}\label{eq:Kt+}
  \begin{aligned}
\mathcal{K}^{t|u,\log}_4=\sum_{i,j,k=0}^\infty\frac{\lambda^2}{128\pi^4}K^{t|u,i,j,k}_{+}u^i(1-v)^j\,,
  \end{aligned}
\end{equation}
in which
\begin{equation}
  \begin{aligned}
K^{t,i,j,k}_{+}=\frac{
4(1+k)\,
\Gamma\!\left(\tfrac{3}{2}+i\right)
\Gamma\!\left(\tfrac{3}{2}+k+i\right)
\Gamma\!\left(\tfrac{3}{2}+i+j\right)
\Gamma\!\left(\tfrac{3}{2}+k+i+j\right)
}{
(1+2k)\,\pi^{2}\,
\Gamma(2+k)\,
\Gamma(1+i)^{2}\,
\Gamma(1+j)\,
\Gamma(3+k+2i+j)
}
\\[6pt]
\times
\left(
-\frac{2}{1+2k}
- H_k
+ H_{\tfrac{1}{2}+k+i}
+ H_{\tfrac{1}{2}+k+i+j}
- H_{2+k+2i+j}
\right)
  \end{aligned}
\end{equation}
and
\begin{equation}\label{eq:Ku+}
  \begin{aligned}
K^{u,i,j,k}_{+}=\frac{
4\,\Gamma\!\left(\tfrac{3}{2}+k+i\right)^{2}
\Gamma\!\left(\tfrac{3}{2}+i+j\right)^{2}
}{
(1+2k)^{2}\,\pi^{2}\,
\Gamma(1+k)\,
\Gamma(1+i)^{2}\,
\Gamma(1+j)\,
\Gamma(3+k+2i+j)
}
\\[6pt]
\times
\left(
-2
+(-1-2k)\,H_k
+(2+4k)\,H_{\tfrac{1}{2}+k+i}
+(-1-2k)\,H_{2+k+2i+j}
\right)\,,
  \end{aligned}
\end{equation}
where $H_{k}$ are the harmonic numbers. It turns out that we may evaluate the infinite sum over $k$ with a clever trick: We define
\begin{equation}
  \begin{aligned}
K^{t,i,j,k}_{-}=\frac{
4(1+k)\,
\Gamma\!\left(\tfrac{3}{2}+i\right)
\Gamma\!\left(\tfrac{3}{2}+k+i\right)
\Gamma\!\left(\tfrac{3}{2}+i+j\right)
\Gamma\!\left(\tfrac{3}{2}+k+i+j\right)
}{
(1+2k)\,\pi^{2}\,
\Gamma(2+k)\,
\Gamma(1+i)^{2}\,
\Gamma(1+j)\,
\Gamma(3+k+2i+j)
}
\\[6pt]
\times
\left(
-\frac{2}{1+2k}
- H_k
+ H_{\tfrac{1}{2}+k+i}
- H_{\tfrac{1}{2}+k+i+j}
+ H_{2+k+2i+j}
\right),
  \end{aligned}
\end{equation}
and
\begin{equation}
  \begin{aligned}
K^{u,i,j,k}_{-}=\frac{
4\,\Gamma\!\left(\tfrac{3}{2}+k+i\right)^{2}
\Gamma\!\left(\tfrac{3}{2}+i+j\right)^{2}
}{
(1+2k)^{2}\,\pi^{2}\,
\Gamma(1+k)\,
\Gamma(1+i)^{2}\,
\Gamma(1+j)\,
\Gamma(3+k+2i+j)
}
\\[6pt]
\times
\left(
-2
-(-1-2k)\,H_k
+(-1-2k)\,H_{2+k+2i+j}
\right),
  \end{aligned}
\end{equation}
which differs from $K^{t,i,j,k}_{+}$, $K^{u,i,j,k}_{+}$ in \eqref{eq:Kt+} and \eqref{eq:Ku+} by a few relative minus signs between the harmonic numbers. Furthermore, we denote $K^{t|u,i,j}_{\pm}=\sum_{k=0}^\infty K^{t|u,i,j,k}_{\pm}$. The evaluation of the infinite sums over $k$ then relies on the following conjectures. 
\begin{conjecture}
    For any $i,j \in \mathbb{Z}$, $K^{t|u,i,j}_{-}$ can be uniquely written as $q_1+\frac{q_2}{\pi}$, in which $q_1,q_2 \in \mathbb{Q}$.
\end{conjecture}
\begin{conjecture}
    $K^{t|u,i,j}_{+} = q_1$.
\end{conjecture}
With the help of Mathematica we were able to compute the infinite sum $K^{t,i,j}_{-}$ and $K^{u,m,n}_{-}$ exactly. It is then verified that $q_1$ is in agreement with the numerical approximation of $K^{t|u,i,j}_{+}$. $K^{t|u,i,j}_{+}$ are then computed in this way on a computer cluster for all $i$ and even $j$ with $i+2j \leq 100$.

$K^{t|u,i,j}_{+}$ with odd $n$ are not helpful to the determination of anomalous dimensions, since they contain information of $(\gamma^{(2)t|u}_{n,l})A_{n,l}G'_{\Delta(n,l),l}$ with $l$ strictly smaller than $n$. That is, they do not provide any new information during the recursive iterations, other than those already known from even $n$ terms. Later on however, after obtaining $\gamma^{(2)t|u}_{n,l}$ from (\ref{AnomDimActualIterationEq}), a consistency check can be performed by plugging $\gamma^{(2)t|u}_{n,l}$ back into the conformal block expansion. This indeed reproduced $K^{t|u,i,j}_{+}$ for odd n as well. It shows that the correlation function correctly admits a conformal block expansion, despite us not being entirely rigorous when we computed the infinite sum over k. 

The $\mathcal{O}(\log u)$ contribution to the conformal block expansion at $\mathcal{O}(\lambda^2)$ comes from

\begin{equation}
\begin{aligned}
    &\left( \mathcal{A}_{n,l}\mathcal{G}_{\Delta(n,l),l} \right)_{\mathcal{O}(\lambda^2),\mathcal{O}(\log u)} \\
    =&\left( \frac{1}{2}(\gamma^{(1)}_{n,l}(\Delta))^22A^{(1)}_{\Delta(n,l),l}G'_{\Delta(n,l),l}+(\gamma^{(2)s}_{n,l}(\Delta))A_{n,l}G'_{\Delta(n,l),l} \right)_{\mathcal{O}(\log u)}.
\end{aligned}
\end{equation}
To evaluate $\gamma^{(2)t|u}_{n,l}(\Delta)$, we manually set $\gamma^{(1)}_{n,l}(\Delta)=0$ during the $t$- and $u$-channel iterations, since $\frac{1}{2}(\gamma^{(1)}_{n,l}(\Delta))^22A^{(1)}_{\Delta(n,l),l}G'_{\Delta(n,l),l}$ has been fully accounted for in the $s$-channel. This leaves us with

\begin{equation}
\begin{aligned}
    W^{\rangle\!\circ\!\langle}_{t|u,\mathcal{O}(\log u)} = \left( (\gamma^{(2)t|u}_{n,l})A_{n,l}G'_{\Delta(n,l),l} \right)_{\mathcal{O}(\log u)}.
    \label{AnomDimActualIterationEq}
\end{aligned}
\end{equation}
Then, $\gamma^{(2)t|u}_{n,l}$ are determined recursively for all even l and $2n+l \leq 100$.
These numbers were fed to the Guess package as part of the RISCergoSUM package \cite{RISCErgoSum, Guess}, with which a recursive relation was found for $\gamma^{(2)t}_{n,l}+\gamma^{(2)u}_{n,l}$, given the initial values. This important result allows us to evaluate the anomalous dimensions for any $n$ and $l$ which is not available in any previous literature to our knowledge. Since the explicit form of this recursion is somewhat lengthy, we  describe it in detail in Appendix \ref{AppendixRecursiveRelation}. With it, the computation time required for $2n+l \approx 100$ was reduced from thousands of hours (that of computing the sum $K^{t,m,n}_{-}$ with Mathematica directly), to merely a few seconds. Some of the first anomalous dimensions are also shown in Table \ref{AnomDimt} and Table \ref{AnomDimu}. Combining them with (\ref{AnomDimS}), we obtained the total contribution from all channels to the anomalous dimensions, some of the first entries shown in Table \ref{AnomDims+t+u}.

\begin{table}[t]
\centering
\resizebox{\textwidth}{!}{$
\begin{array}{c|ccccc}
\gamma_{n,l}^{t} & l=0 & l=2 & l=4 & l=6 & l=8 \\
\noalign{\vskip 0.6ex}
\hline
\noalign{\vskip 0.6ex}
n=0 &
-\dfrac{\lambda^2}{192\pi^2} &
\dfrac{5\lambda^2}{4032\pi^2} &
-\dfrac{299\lambda^2}{24640\pi^2} &
\dfrac{68861\lambda^2}{172480\pi^2} &
-\dfrac{347063483\lambda^2}{12640320\pi^2}
\\[1.2em]
n=1 &
-\dfrac{\lambda^2}{5040\pi^2} &
-\dfrac{4777\lambda^2}{221760\pi^2} &
\dfrac{710383\lambda^2}{720720\pi^2} &
-\dfrac{160778091931\lambda^2}{1862340480\pi^2} &
\dfrac{29966280403819\lambda^2}{2433740400\pi^2}
\\[1.2em]
n=2 &
-\dfrac{811973\lambda^2}{29937600\pi^2} &
\dfrac{4932359\lambda^2}{2882880\pi^2} &
-\dfrac{19107541407709\lambda^2}{97772875200\pi^2} &
\dfrac{59710558672625843\lambda^2}{1734770157120\pi^2} &
-\dfrac{91522523466052657037\lambda^2}{10601373182400\pi^2}
\\[1.2em]
n=3 &
\dfrac{20709259\lambda^2}{10090080\pi^2} &
-\dfrac{110254681709\lambda^2}{349188840\pi^2} &
\dfrac{1503137554547357\lambda^2}{21416915520\pi^2} &
-\dfrac{29911835316920716043\lambda^2}{1405485081000\pi^2} &
\dfrac{222834973307634513509129\lambda^2}{26369344441440\pi^2}
\\[1.2em]
n=4 &
-\dfrac{2711482461412687\lambda^2}{7332965640000\pi^2} &
\dfrac{5035734323168371943\lambda^2}{47224298721600\pi^2} &
-\dfrac{237059057696361975071\lambda^2}{5952642696000\pi^2} &
\dfrac{6271272444738145656083221381\lambda^2}{333572207184216000\pi^2} &
-\dfrac{333023559517979601428378445023\lambda^2}{30324746107656000\pi^2}
\end{array}
$}
\caption{Some lowest order $\gamma^{(2),t}_{n,l}$}
\label{AnomDimt}
\end{table}

\begin{table}[t]
\centering
\resizebox{\textwidth}{!}{$
\begin{array}{c|ccccc}
\gamma_{n,l}^{u} & l=0 & l=2 & l=4 & l=6 & l=8 \\
\noalign{\vskip 0.6ex}
\hline
\noalign{\vskip 0.6ex}
n=0 &
-\dfrac{\lambda^2}{192\pi^2} &
-\dfrac{31\lambda^2}{20160\pi^2} &
\dfrac{2677\lambda^2}{221760\pi^2} &
-\dfrac{2685733\lambda^2}{6726720\pi^2} &
\dfrac{5900076901\lambda^2}{214885440\pi^2}
\\[1.2em]
n=1 &
-\dfrac{37\lambda^2}{10080\pi^2} &
\dfrac{131\lambda^2}{6160\pi^2} &
-\dfrac{4262627\lambda^2}{4324320\pi^2} &
\dfrac{10048627093\lambda^2}{116396280\pi^2} &
-\dfrac{59932560885053\lambda^2}{4867480800\pi^2}
\\[1.2em]
n=2 &
\dfrac{750203\lambda^2}{29937600\pi^2} &
-\dfrac{10570753\lambda^2}{6177600\pi^2} &
\dfrac{6369178024393\lambda^2}{32590958400\pi^2} &
-\dfrac{59710558731540149\lambda^2}{1734770157120\pi^2} &
\dfrac{18304504693171821479\lambda^2}{2120274636480\pi^2}
\\[1.2em]
n=3 &
-\dfrac{2072237\lambda^2}{1009008\pi^2} &
\dfrac{73503077213\lambda^2}{232792560\pi^2} &
-\dfrac{751568778021689\lambda^2}{10708457760\pi^2} &
\dfrac{59823670633746001289\lambda^2}{2810970162000\pi^2} &
-\dfrac{4271003655063004511982203\lambda^2}{505412435127600\pi^2}
\\[1.2em]
n=4 &
\dfrac{2711475855567217\lambda^2}{7332965640000\pi^2} &
-\dfrac{1007146866120317821\lambda^2}{9444859744320\pi^2} &
\dfrac{16584378521941152371\lambda^2}{416440024000\pi^2} &
-\dfrac{6271272444738156592681917019\lambda^2}{333572207184216000\pi^2} &
\dfrac{333023559517979600845938497969\lambda^2}{30324746107656000\pi^2}
\end{array}
$}
\caption{Some lowest order $\gamma^{(2),u}_{n,l}$}
\label{AnomDimu}
\end{table}

\begin{table}[t]
\centering
\resizebox{\textwidth}{!}{$
\begin{array}{c|ccccc}
\gamma_{n,l} & l=0 & l=2 & l=4 & l=6 & l=8 \\
\noalign{\vskip 0.6ex}
\hline
\noalign{\vskip 0.6ex}
n=0 &
-\dfrac{11\lambda^2}{768\pi^2} &
-\dfrac{\lambda^2}{3360\pi^2} &
-\dfrac{\lambda^2}{15840\pi^2} &
-\dfrac{\lambda^2}{43680\pi^2} &
-\dfrac{\lambda^2}{93024\pi^2}
\\[1.2em]
n=1 &
-\dfrac{937\lambda^2}{215040\pi^2} &
-\dfrac{61\lambda^2}{221760\pi^2} &
-\dfrac{47\lambda^2}{617760\pi^2} &
-\dfrac{253\lambda^2}{8062080\pi^2} &
-\dfrac{397\lambda^2}{24961440\pi^2}
\\[1.2em]
n=2 &
-\dfrac{17627\lambda^2}{7983360\pi^2} &
-\dfrac{4943\lambda^2}{21621600\pi^2} &
-\dfrac{104779\lambda^2}{1396755360\pi^2} &
-\dfrac{297547\lambda^2}{8761465440\pi^2} &
-\dfrac{676747\lambda^2}{37067738400\pi^2}
\\[1.2em]
n=3 &
-\dfrac{1004021\lambda^2}{738017280\pi^2} &
-\dfrac{131779\lambda^2}{698377680\pi^2} &
-\dfrac{1496021\lambda^2}{21416915520\pi^2} &
-\dfrac{1239361\lambda^2}{36506106000\pi^2} &
-\dfrac{57960383\lambda^2}{3029445165600\pi^2}
\\[1.2em]
n=4 &
-\dfrac{86793659\lambda^2}{93117024000\pi^2} &
-\dfrac{58993787\lambda^2}{374796021600\pi^2} &
-\dfrac{460967861\lambda^2}{7228208988000\pi^2} &
-\dfrac{23672291549\lambda^2}{722017764468000\pi^2} &
-\dfrac{13867617787\lambda^2}{722017764468000\pi^2}
\end{array}
$}
\caption{Some lowest order $\gamma^{(2),s+t+u}_{n,l}$}
\label{AnomDims+t+u}
\end{table}

An interesting feature that emerged in our analysis is a fine-tuning-like behavior in the $t$- and $u$-channel anomalous dimensions. In particular, at large $2n+l$ limit, the combined contribution $\gamma^{(2),t+u}_{n,l}$ is magnitudes of order smaller than the individual contributions $\gamma^{(2),t}_{n,l}$ and $\gamma^{(2),u}_{n,l}$ (at $2n+l \sim 100$, this magnitude could be up to $10^{200}$). While (\ref{AnomDimSpliting}) introduces redundant degrees of freedom in the separate definitions of $\gamma^{(2),t}_{n,l}$ and $\gamma^{(2),u}_{n,l}$ such that only their sum represents a physical quantity, we nevertheless speculate that this 'fine-tuning' may be related to an underlying symmetry shared by the two channels. It would be interesting to investigate whether this behavior admits an interpretation in terms of the $6j$ symbols. We leave a detailed exploration of this possibility to future work.

In addition, for $n=0$ we were able to guess a closed form for the anomalous dimensions as a function of $l$:
\begin{equation}
  \begin{aligned}
    \gamma^{s+t+u}_{0,l}=-\frac{\lambda^2}{32\pi^2 (l+1)(2l+1)(2l+3)}-\frac{\lambda^2}{256\pi^2}\delta^{l0}.
  \end{aligned}
\end{equation}
This is another key result of this paper. 
If we substitute the intrinsic spin $l$ with the conformal spin $J$, defined as
\begin{equation}
  \begin{aligned}
    J^2=(l+\tau/2)(l+\tau/2-1)=(l+\frac{3}{2})(l+\frac{1}{2})\,,
  \end{aligned}
\end{equation}
where $\tau=2\Delta=3$, the anomalous dimensions become
\begin{equation}
  \begin{aligned}
    \gamma^{s+t+u}_{0,l}&=\frac{\lambda^2}{64\pi^4 J^2 \sqrt{1+4J^2}} \\
    &=\frac{\lambda^2}{128\,\pi^{4}}
\left(
\frac{1}{J^{3}}
-\frac{1}{8 J^{5}}
+\frac{3}{128 J^{7}}
-\frac{5}{1024 J^{9}}
+\frac{35}{32768 J^{11}}
+\mathcal{O}\!\left(J^{-13}\right)
\right)
  \end{aligned}
\end{equation}
when expanded at large $J$ limit. The series in J contains only $\frac{1}{J^\tau}$ times even power terms of $J$, which agrees with \cite{Basso:2006nk,Alday:2015eya,Alday:2015ewa,Bissi:2022mrs}.
To summarize, using a new representation of one-loop diagrams in terms of an infinite sum of tree level diagrams, combined with some intricate manipulations of these infinite sums inside the conformal block expansion, we were able to extract a recursion relation for all second order anomalous dimensions of the deformed 2-dimensional conformal field theory corresponding to an interacting scalar field theory in $\text{AdS}_3$.

\section{Conclusion}
In this paper, we computed the four-point one-loop correlator of a conformally coupled $\phi^4$ interacting scalar field in $\text{AdS}_3$. Direct evaluation of the associated Feynman integrals proved challenging due to the non-integer powers arising from the conformal dimension of the field. However, we found that the one-loop correlator admits a series expansion in tree-level 'cross' diagrams with progressively increasing conformal dimensions in the bulk-to-boundary propagators. 

In the $s$-channel, this discovery allows the spectral function of the loop diagram to be expressed as a sum of cross-diagram spectral functions. From this spectral function, we extracted the anomalous dimensions and OPE coefficients dual to the bulk one-loop corrections, finding agreement with a different representation of the spectral function obtained by bootstrap. In contrast, this approach fails in the $t$- and $u$-channels due to a mismatch in the conformal partial waves. To overcome this obstacle, we instead performed conformal block expansion for each cross diagram appearing in the series. The infinite sum of cross diagrams was evaluated under the assumption of two conjectures proposed in this work, which allowed us to obtain exact values of the anomalous dimensions in the $t$- and $u$-channels. Moreover, we identified a recursive relation that enables the efficient computation of anomalous dimensions at arbitrarily high orders given a few initial values. 

Looking ahead, it would be interesting to extend this analysis to the alternative conformally coupled scalar with $\Delta = \tfrac{1}{2}$ and explore possible connections to known two-dimensional conformal field theories. Another promising direction is to investigate potential links between the framework developed here and recent approaches to cosmological correlators.

\acknowledgments
 W.X. would like to thank the China Scholarship Council for sponsoring his doctoral study. W.X. would also like to thank Daniel Bockisch, Eugenia Boffo, Dean Carmi, Ka Hei Choi, Riccardo Ciccone, Carlo Alberto Cremonini, Zhen Gao, Jonathan Gräfe, Matthias Nowinski, Shimon Sukholuski and Dayuan Wang for discussions. Specifically, W.X. would like to thank Till Heckelbacher for supervision of his master project, from which the present work is developed.
This work is supported by the Excellence Cluster Origins of the DFG under Germany's Excellence Strategy EXC-2094 390783311 as well as EXC 2094/2: ORIGINS 2. 

\appendix
\section{Loop calculation}

\label{AppendixJ4integraldetail}

We discuss in detail the calculation of the fish diagram
\begin{equation}
\begin{aligned}
    \mathcal{J}_4=\int \text{d}^3y \sqrt{|g_y|} \Lambda(x,y)^2 \overline{\Lambda}(y,x_3) \overline{\Lambda}(y,x_4),
\end{aligned}
\end{equation}
shown in Figure \ref{fig:J4}. Similar to \cite{Bertan:2018afl}, the following isometries are applied to simplify the integration: 
\begin{enumerate}
    \item Translation of $x,x_3,x_4$ by $(-x_4^i,0)$, denoted by the first prime, such that $x_4'=(0,0)$. 
    \item Inversion of $x',y,x_3',x_4'$, denoted by the second prime, such that
\begin{equation}
K_{x'y}=K_{x''y''}\quad \overline{K}_{x_3'y}=\frac{1}{{x_3'}^2} \overline{K}_{x_3''y''}=\frac{1}{r_{34}^2}\overline{K}_{x_3''y''}\quad \overline{K}_{x_4'y}={x_4''}^2\overline{K}_{x_4''y''}=2w''\,.
\label{isometriesJ4}
\end{equation}
\item Translation $(y^i,w)=(y''^i-x''^i,w'')$, as well as denoting $x_3'''^i=x_3''^i-x''^i$.
\end{enumerate}
The fish diagram then reads
\begin{equation}
  \begin{aligned}
\mathcal{J}_4=&\int \text{d}^3y \frac{a}{128\pi^4} \frac{1}{r_{34}^3} (z_3 z_4)^\frac{3}{2} (\frac{1}{(y-x_3''')^2+w^2})^\frac{3}{2} \bigg( \frac{2wz''}{y^2+(w-z'')^2} \\
&+ \frac{2wz''}{y^2+(w+z'')^2} - 2\sqrt{\frac{2wz''}{y^2+(w-z)^2}\frac{2wz}{y^2+(w+z'')^2}} \bigg) \\
\defeq&\mathcal{J}_4^1+\mathcal{J}_4^2-\mathcal{J}_4^3,
 \end{aligned}
\end{equation}
where the three terms are to be computed separately.
\begin{equation}
  \begin{aligned}
\mathcal{J}_4^1=&\int \text{d}^3y \frac{a}{128\pi^4} \frac{1}{r_{34}^3} (z_3 z_4)^\frac{3}{2} (\frac{1}{(y-x_3''')^2+w^2})^\frac{3}{2} \frac{2wz''}{y^2+(w-z'')^2}\\
=&\int \text{d}^3y C (\frac{1}{(y-x_3)^2+w^2})^\frac{3}{2} \frac{w}{y^2+(w-z'')^2} \,,
 \end{aligned}
\end{equation}
in which $C=\frac{az''}{64\pi^4} \frac{1}{r_{34}^3} (z_3 z_4)^\frac{3}{2}$. Next we make use of the Schwinger parametrization to get
\begin{equation}
  \begin{aligned}
\mathcal{J}_4^1=&\int_0^{\infty} \text{d}w \text{d}t_1 \text{d}t_2 C \frac{1}{\Gamma(\frac{3}{2})} \frac{1}{t_1+t_2} \pi  \sqrt{t_1} w e^{-\frac{t_1^2 w^2+t_2^2 (w-{z''})^2+t_1 t_2 \left(2 w^2+x_3'''^2-2
w {z''}+{z''}^2\right)}{t_1+t_2}}\,.
 \end{aligned}
\end{equation}
A change of variable is then made substituting $t_i=ss_i$, and $\text{d}t_1\text{d}t_2=s\text{d}s\text{d}s_1\text{d}s_2$. The integration is then performed in the order of $s \rightarrow w \rightarrow s_2$,
\begin{equation}
  \begin{aligned}
\mathcal{J}_4^1=&\int \text{d}w \text{d}s \text{d}s_1 ds_2 C \frac{1}{\Gamma(\frac{3}{2})} \delta(1-s_1-s_2) \pi  s \sqrt{s_1*s} w \\
&e^{-s\left(s_1^2 w^2+s_2^2 (w-{z''})^2+s_1 s_2 \left(2 w^2+x_3'''^2-2 w {z''}+{z''}^2\right)\right)} \\
=&\int_0^1 \text{d}s_1 C \frac{1}{\Gamma(\frac{3}{2})} \frac{\pi ^{3/2} \left({z''}-s_1 {z''}+\sqrt{(1-s_1) \left(s_1 x_3'''^2+{z''}^2\right)}\right)}{2
(1-s_1) \sqrt{s_1} \left(x_3'''^2+{z''}^2\right)}\,.
 \end{aligned}
\end{equation}
Similarly,
\begin{equation}
  \begin{aligned}
\mathcal{J}_4^2=&\int_0^1 \text{d}s_1 C \frac{1}{\Gamma(\frac{3}{2})} \frac{\pi ^{3/2} \left(s_1 {z''}-{z''}+\sqrt{(1-s_1) \left(s_1 x_3'''^2+{z''}^2\right)}\right)}{2
(1-s_1) \sqrt{s_1} x_3'''^2},
 \end{aligned}
\end{equation}
so that
\begin{equation}
  \begin{aligned}
\mathcal{J}_4^1+\mathcal{J}_4^2=&\int_0^1 \text{d}t C \frac{1}{\Gamma(\frac{3}{2})} \frac{2\pi ^{3/2} \left(\sqrt{(1-t^2) \left(t^2 \left(x_3'''^2\right)+{z''}^2\right)}\right)}{
(1-t^2) x_3'''^2} \\
=&C \frac{1}{\Gamma(\frac{3}{2})} \frac{2\pi ^{3/2} {z''} \text{EllipticE}[-\frac{x_3'''^2}{{z''}^2}]}{
x_3'''^2+{z''}^2}.
 \end{aligned}
\end{equation}
Finally, 
\begin{equation}
  \begin{aligned}
\mathcal{J}_4^3=&2\int \text{d}^3y \frac{a}{128\pi^4} \frac{1}{r_{34}^3} (z_3 z_4)^\frac{3}{2} (\frac{1}{(y-x_3''')^2+w^2})^\frac{3}{2} \sqrt{\frac{2wz''}{y^2+(w-z'')^2}} \sqrt{\frac{2wz''}{y^2+(w+z'')^2}}\\
=&2\int \text{d}w \text{d}t_1 \text{d}t_2 \text{d}t_3 C \frac{1}{\Gamma(\frac{3}{2})} \frac{1}{\Gamma(\frac{1}{2})} \frac{1}{\Gamma(\frac{1}{2})} \frac{
\pi  \sqrt{t_1} w}{\sqrt{t_2} \sqrt{t_3} (t_1+t_2+t_3)} \\
&e^{-\frac{t_1^2 w^2+t_2^2 (w-z'')^2+t_3^2 (w+z'')^2+2 t_2 t_3 \left(w^2+z''^2\right)+t_1
\left(t_2 \left(2 w^2+x_3'''^2-2 w z''+z''^2\right)+t_3 \left(2 w^2+x_3'''^2+2 w z''+z''^2\right)\right)}{t_1+t_2+t_3}}.
 \end{aligned}
\end{equation}
Substituting $t_i=ss_i$, $\text{d}t_1\text{d}t_2\text{d}t_3=s^2\text{d}s\text{d}s_1\text{d}s_2\text{d}s_3$, and performing integration in the order of $s \rightarrow w \rightarrow s_1$,
\begin{equation}
  \begin{aligned}
\mathcal{J}_4^3=&2\int \text{d}w \text{d}s \text{d}s_1 \text{d}s_2 \text{d}s_3 C \frac{1}{\Gamma(\frac{3}{2})} \frac{1}{\Gamma(\frac{1}{2})} \frac{1}{\Gamma(\frac{1}{2})} \frac{\pi \sqrt{s s_1} w}{\sqrt{s_2} \sqrt{s_3}} \delta(1-s_1-s_2-s_3)\\
&e^{-s\left(s_1^2 w^2+s_1 \left((s_2+s_3) \left(2 w^2+x_3'''^2\right)+2
(-s_2+s_3) w z''+(s_2+s_3) z''^2\right)+(s_2+s_3) \left(s_2 (w-z'')^2+s_3 (w+z'')^2\right)\right)} \\
=&2\int^{0\leq s_2+s_3\leq 1} ds_2  ds_3 C \frac{1}{\Gamma(\frac{3}{2})} \frac{1}{\Gamma(\frac{1}{2})} \frac{1}{\Gamma(\frac{1}{2})} \pi ^{3/2} \sqrt{1-s_2-s_3} \\
&\frac{(s_2-s_3) z''+\sqrt{(s_2+s_3) \left((1-s_2-s_3)
x_3'''^2+z''^2\right)}}{2 \sqrt{s_2} \sqrt{s_3} \left(4 s_2 s_3 z''^2+(1-s_2-s_3)
(s_2+s_3) \left(x_3'''^2+z''^2\right)\right)}.
 \end{aligned}
\end{equation}
Changing the integration parameters into polar coordinates,
\begin{equation}
  \begin{aligned}
\mathcal{J}_4^3=&2\int^{0\leq a^2+b^2\leq 1} 4 \text{d}a \text{d}b C \frac{1}{\Gamma(\frac{3}{2})} \frac{1}{\Gamma(\frac{1}{2})} \frac{1}{\Gamma(\frac{1}{2})} \\
&\frac{\sqrt{1-a^2-b^2} \pi ^{3/2} \left(\left(a^2-b^2\right) z''+\sqrt{\left(a^2+b^2\right) \left(\left(1-a^2-b^2\right) x_3'''^2+z''^2\right)}\right)}{2 \left(4
a^2 b^2 z''^2+\left(1-a^2-b^2\right) \left(a^2+b^2\right) \left(x_3'''^2+z''^2\right)\right)} \\
=&2\int_0^1 4 r \text{d}r \int_0^{\pi/2} \text{d}\theta C \frac{1}{\Gamma(\frac{3}{2})} \frac{1}{\Gamma(\frac{1}{2})} \frac{1}{\Gamma(\frac{1}{2})} \\
&\frac{\pi ^{3/2} \sqrt{1-r^2} \left(\sqrt{r^2 \left(\left(1-r^2\right) x_3'''^2+z''^2\right)}+r^2 z'' \text{cos}(2
\theta )\right)}{2 \left(r^2 \left(1-r^2\right) \left(x_3'''^2+z''^2\right)+r^4 z''^2 \text{sin}(2 \theta )^2 \right)}.
 \end{aligned}
\end{equation}
We can finally solve the integral in the order of $\theta \rightarrow r$
\begin{equation}
  \begin{aligned}
\mathcal{J}_4^3=&2C \frac{1}{\Gamma(\frac{3}{2})} \frac{1}{\Gamma(\frac{1}{2})} \frac{1}{\Gamma(\frac{1}{2})} \pi^{5/2} \sqrt{\frac{1}{x_3'''^2+z''^2}}.
 \end{aligned}
\end{equation}
Summing the three terms up,
\begin{equation}
  \begin{aligned}
\mathcal{J}_4 =&\mathcal{J}_4^1+\mathcal{J}_4^2-\mathcal{J}_4^3 \\
=&\frac{a}{16\pi^3} \frac{1}{r_{34}^3} (z_3 z_4)^\frac{3}{2} \left(\frac{{z''}^2 \text{EllipticE}[-\frac{x_3'''^2}{{z''}^2}]}{x_3'''^2+{z''}^2} - \sqrt{\frac{{z''}^2}{x_3'''^2+{z''}^2}}\right) \\
=&\frac{a \lambda}{128\pi^3} (z_3 z_4)^\frac{3}{2} (\overline{K}_{x x_3}  \overline{K}_{x x_4})^\frac{3}{2} \left( \sqrt{1+\alpha^2}\text{EllipticE}[-\alpha^2] - (1+\alpha^2) \right)\,,
 \end{aligned}
\end{equation}
where the covariant quantity $\alpha$ is defined in (\ref{recovercovariance}).

\section{Spectral Function}
\label{AppendixSpectralFunction}
In this section we recap how to extract dual CFT data from spectral functions of bulk correlator. 
\subsection{Disconnected Diagrams}
The spectral function of the disconnected diagram, given in \cite{Sachs:2023eph} for $\text{AdS}_4$ and $\Delta=2$, can be generalized to any spacetime and conformal dimension as
\begin{equation}
  \begin{aligned}
D^{(\text{disc})}(\nu)=\frac{\Gamma(\frac{d}{2}-\Delta)^2\Gamma(\Delta-\frac{d}{4}+i\frac{\nu}{2})\Gamma(-\frac{d}{4}+\Delta-i\frac{\nu}{2})\Gamma(\frac{d}{2})\Gamma(\frac{d}{2}-i\nu)\Gamma(\frac{d}{4}+i\frac{\nu}{2})^2}{2\pi\Gamma(\frac{3d}{4}-\Delta+i\frac{\nu}{2})\Gamma(\frac{3d}{4}-\Delta-i\frac{\nu}{2})\Gamma(i\nu)\Gamma(\frac{d}{4}-i\frac{\nu}{2})^2\Gamma(\Delta)^2},
\label{Ddisc}
  \end{aligned}
\end{equation}
with the unperturbed, $0$-th order OPE coefficients given by
\begin{equation}
  \begin{aligned}
A_{n,0}=-2\pi i\text{Res}D^{(\text{disc})}(\nu_n).
\label{OPEdisc}
  \end{aligned}
\end{equation}
\subsection{Tree Level Contact Interaction}
\begin{figure}[t]
\begin{tikzpicture}
  \tikzfeynmanset{
    every vertex={black, dot},
  }
  \begin{feynman}
    \vertex (center) {};
    \vertex [left=0.8cm of center] (a);
    \vertex [right=0.8cm of center] (b);
    \vertex [above=1cm of center] (x0);
    \vertex [left=0.8cm of a] (atext) {\(Y_1\)};
    \vertex [right=0.8cm of b] (btext) {\(Y_2\)};
    \vertex [above=0.4cm of x0] (x0text) {\(x_0\)};
    \vertex [above left=of a] (f1);
    \vertex [left=0.8cm of f1] (f1text) {\(x_1\)};
    \vertex [below left=of a] (f2);
    \vertex [left=0.8cm of f2] (f2text) {\(x_2\)};
    \vertex [above right=of b] (f3);
    \vertex [right=0.8cm of f3] (f3text) {\(x_3\)};
    \vertex [below right=of b] (f4);
    \vertex [right=0.8cm of f4] (f4text) {\(x_4\)};

    \diagram* {
      (a) -- [plain] (f1),
      (a) -- [plain] (f2),
      (b) -- [plain] (f3),
      (b) -- [plain] (f4),
      (a) -- [dashed] (x0),
      (b) -- [dashed] (x0),
    };
  \end{feynman}
\end{tikzpicture}
\centering
\caption{Spectral representation of the cross diagram}
\label{fig:I4SpectralRep}
\end{figure}
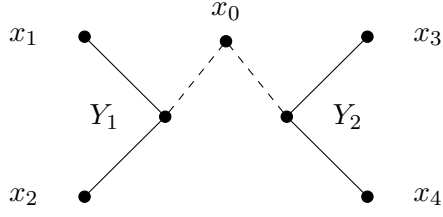

We start with the boundary three point function given by \cite{Meltzer:2019nbs} 
\begin{equation}
  \begin{aligned}
W_\nu(x_1,x_2,x_0) =& \int_{\text{AdS}_{d+1}} \text{d}^{d+1}y \overline{\Lambda}_\Delta(x_1,y)\overline{\Lambda}_\Delta(x_2,y)\overline{\Lambda}_{\Delta(\nu)}(x_0,y) \\
=& b_{\Delta,\Delta,\nu}\frac{1}{x_{12}^{2\Delta-\Delta(\nu)}x_{10}^{\Delta(\nu)}x_{20}^{\Delta(\nu)}},
  \end{aligned}
\end{equation}
\begin{equation}
  \begin{aligned}
b_{\Delta,\Delta,\nu} =& \frac{1}{2^4\pi^{d}}\frac{\Gamma(\Delta-\frac{d}{4}+i\frac{\nu}{2}) \Gamma(\Delta-\frac{d}{4}-i\frac{\nu}{2})\Gamma(\frac{d}{4}+i\frac{\nu}{2})^2}{\Gamma(\Delta+1-\frac{d}{2})^2\Gamma(1+i\nu)},
    \label{threepointfunction}
  \end{aligned}
\end{equation}
out of which we the reconstruct the 4-point interaction by insertion of a delta function in the bulk represented as
\begin{equation}
  \begin{aligned}
\delta(Y_1-Y_2)=\int_\mathbb{R} \text{d}\nu \frac{\nu^2}{\pi} \int_{\text{AdS}_{d+1}} \text{d}^d x_0 \overline{\Lambda}_{\Delta(\nu)}(Y_1,x_0)\overline{\Lambda}_{\Delta(-\nu)}(Y_2,x_0)
\label{deltafunction}
  \end{aligned}
\end{equation}
as shown in Figure \ref{fig:I4SpectralRep}. The four point, tree level diagram is then reconstructed as 
\begin{equation}
  \begin{aligned}
W^{\times}(x_1,x_2,x_3,x_4)=& -\lambda\int_\mathbb{R} \text{d}\nu \frac{\nu^2}{\pi} \int_{\partial AdS_{d+1}} \text{d}^{d}x_0 W_\nu(x_1,x_2,x_0) W_{-\nu}(x_0,x_3,x_4) \\
=& -\lambda \int_\mathbb{R} \text{d}\nu \frac{\nu^2}{\pi} b_{\Delta,\Delta,\nu} b_{\Delta,\Delta,-\nu} \Psi^{\Delta,\Delta,\Delta,\Delta}_{x_1,x_2,x_3,x_4,\nu} \\
=& -\lambda \int_\mathbb{R} \text{d}\nu \frac{\nu^2}{\pi} b_{\Delta,\Delta,\nu} b_{\Delta,\Delta,-\nu} K(\nu) g^{\Delta,\Delta,\Delta,\Delta}_{x_1,x_2,x_3,x_4,\nu} \\
:=& \int_\mathbb{R} \text{d}\nu D^\times(\nu) g^{\Delta,\Delta,\Delta,\Delta}_{x_1,x_2,x_3,x_4,\nu}.
  \end{aligned}
  \label{CrossDiagramAsCPW}
\end{equation}
where we have used (\ref{eq:WDg}-\ref{eq:K}). Hence we read off
\begin{equation}
  \begin{aligned}
D^{\times}(\nu) = -\lambda \frac{\nu^2}{\pi} b_{\Delta,\Delta,\nu} b_{\Delta,\Delta,-\nu} K(\nu).
  \end{aligned}
\end{equation}
The conformal block $g^{\Delta,\Delta,\Delta,\Delta}_{x_1,x_2,x_3,x_4,\nu}$ is regular while $D^{cross}(\nu)$ has double poles at $\nu_n=\Delta+n-\frac{d}{4}$. Therefore, after closing the contour in the $-\nu$ half plane, residue theorem yields
\begin{equation}
  \begin{aligned}
&W^{\times}(x_1,x_2,x_3,x_4)\\
=& -2\pi i \sum_{n=0}^\infty \text{Res} \left( D^{\times}(\nu) g^{\Delta,\Delta,\Delta,\Delta}_{x_1,x_2,x_3,x_4,\nu}, \nu=\nu_n \right) \\
=&\sum_{n=0}^\infty \left( \underbrace{-2\pi iD^{\times^{(-2)}}_{\nu_n} g^{(1)}(x_1,x_2,x_3,x_4,\nu_n)}_{=\gamma^{(1)}_{n,0}A_{n,0}g'_{n,0}} - \underbrace{2\pi iD^{\times^{(-1)}}_{\nu_n} g^{(0)}(x_1,x_2,x_3,x_4,\nu_n)}_{=\gamma^{(1)}_{n,0} A^{(1)}_{n,0}g_{n,0}} \right).
\label{crossresidue}
  \end{aligned}
\end{equation}
where $f^{(i)}_{\nu_n}$ denotes the i-th order Laurent series coefficient of $f(\nu)$ expanded around $\nu=\nu_n$, and in the last line a comparison is made to the conformal block expansion of the cross diagram,
\begin{equation}
  \begin{aligned}
W^{\times}(x_1,x_2,x_3,x_4) = \sum_{n=0}^\infty \left( \gamma^{(1)}_{n,0}A_{n,0}g'_{n,0} + \gamma^{(1)}_{n,0} A^{(1)}_{n,0}g_{n,0} \right).
\label{crossconformalblock}
  \end{aligned}
\end{equation}
Plugging (\ref{OPEdisc}) into (\ref{crossresidue}), we conclude that
\begin{equation}
  \begin{aligned}
\gamma^{(1)}_{n,0}=\frac{-2\pi iD^{\times^{(-2)}}_{\nu_n}}{A_{n,0}}=\frac{D^{\times^{(-2)}}_{\nu_n}}{\text{Res}D^{(\text{disc})}(\nu_n)},
\label{Dcross-2}
  \end{aligned}
\end{equation}
\begin{equation}
  \begin{aligned}
A^{(1)}_{n,0}=\frac{-2\pi iD^{\times^{(-1)}}_{\nu_n}}{\gamma^{(1)}_{n,0}}.
\label{OPEcoeff1storder}
  \end{aligned}
\end{equation}
In the special case of $d=2$ and $\Delta=\frac{3}{2}$, we have 
\begin{equation}
  \begin{aligned}
\gamma^{(1)}_{n,0}=\frac{\lambda}{16\pi(n+1)}
  \end{aligned}
\end{equation}
and it is verified that (\ref{OPEcoeff1storder}) agrees with the formula given by \cite{Fitzpatrick:2011dm} that
\begin{equation}
  \begin{aligned}
A^{(1)}_{n,0}=\frac{1}{2 \gamma^{(1)}_{n,0}}\frac{\partial(A_{n,0}\gamma^{(1)}_{n,l})}{\partial n},
\label{OPEcoeff1storderFitzpatrickKaplan}
  \end{aligned}
\end{equation}
in which our conventions are related to \cite{Fitzpatrick:2011dm} by
\begin{equation}
  \begin{aligned}
A_{n,l}+\gamma^{(1)}_{n,l}A^{(1)}_{n,l}=(\overline{c}_{n,l}+\delta c_{n,l})^2, \quad A_{n,l}=\overline{c}_{n,l}^2, \quad \gamma^{(1)}_{n,l}A^{(1)}_{n,l}=2\overline{c}_{n,l}\delta c_{n,l}.
  \end{aligned}
\end{equation}

\subsection{one-loop Level}

It has been observed in \cite[3.1]{Sachs:2023eph} that one can make use of the delta function (\ref{deltafunction}) to write the one-loop level spectral function for the $s$-channel as
\begin{equation}
  \begin{aligned}
D^{\rangle\!\circ\!\langle}(\nu) = D^{\times}(\nu) B(\nu),
\label{K-Lspecrep}
  \end{aligned}
\end{equation}
where $B(\nu)$ is the Källén–Lehmann spectral representation of the intermediate double trace operator that is dual to the bubble diagram (considered a two point function in the $x_1 \rightarrow x_2$ and $x_3 \rightarrow x_4$ limit), as shown in Figure \ref{DoubleTraceSpecRep}. The $s$-channel four-point function at one-loop level thus reads
\begin{equation}
  \begin{aligned}
W^{\rangle\!\circ\!\langle}(x_1,x_2,x_3,x_4)=& \int_\mathbb{R} \frac{\text{d}\nu}{2\pi} D^{\times}(\nu) B(\nu) g^{\Delta,\Delta,\Delta,\Delta}_{x_1,x_2,x_3,x_4,\nu} \\
=&2\pi i \text{Res}\left(\frac{1}{2\pi} D^{\times}(\nu) B(\nu) g^{\Delta,\Delta,\Delta,\Delta}_{x_1,x_2,x_3,x_4,\nu}, \nu=\nu_n \right).
  \end{aligned}
\end{equation}
We expect $B(\nu)$ to have simple poles at $\nu_n=\Delta+n-\frac{d}{4}$ \footnote{This is verified by (\ref{spectralfunctionfromellipticexpansion}).}, resulting in the following expansion
\begin{equation}
  \begin{aligned}
&2\pi i\text{Res} \left( D^{\times}(\nu) B(\nu) g^{\Delta,\Delta,\Delta,\Delta}_{x_1,x_2,x_3,x_4,\nu},\nu=\nu_n \right) \\
=& \underbrace{2\pi iD^{\times^{(-2)}}_{\nu_n} B^{(-1)}_{\nu_n} g^{(2)}(\nu_n)}_{=\frac{1}{2}(\gamma^{(1)}_{n,0})^2A_{n,0}g''_{n,0}} + \underbrace{2\pi iD^{\times^{(-1)}}_{\nu_n} B^{(-1)}_{\nu_n} g^{(1)}(\nu_n)}_{=(\gamma^{(1)}_{n,0})^2A_{n,0}^{(1)}g'_{n,0}} + \underbrace{2\pi iD^{\times^{(-2)}}_{\nu_n} B^{(0)}_{\nu_n} g^{(1)}(\nu_n)}_{=\gamma^{(2)}_{n,0}A_{n,0} g'_{n,0}} \\
+& \underbrace{2\pi iD^{\times^{(0)}}_{\nu_n} B^{(-1)}_{\nu_n} g^{(0)}(\nu_n) + 2\pi iD^{\times^{(-2)}}_{\nu_n} B^{(1)}_{\nu_n} g^{(0)}(\nu_n)}_{=\frac{1}{2}(\gamma^{(1)}_{n,0})^2A_{n,0}^{(2)}g_{n,0}} + \underbrace{2\pi iD^{\times^{(-1)}}_{\nu_n} B^{(0)}_{\nu_n} g^{(0)}(\nu_n)}_{=\gamma^{(2)}_{n,0}A_{n,0}^{(1)} g_{n,0}}, \\
\label{oneloopspecfunctionres}
  \end{aligned}
\end{equation}
where each term is matched with their counterpart in the conformal block expansion,
\begin{equation}
  \begin{aligned}
&W^{\rangle\!\circ\!\langle}(x_1,x_2,x_3,x_4) \\
=& \sum_{n=0}^\infty \left( \frac{1}{2} (\gamma^{(1)}_{n,0})^2 \left( A_{n,0}g''_{n,0} + 2A_{n,0}^{(1)}g'_{n,0} + A_{n,0}^{(2)}g_{n,0} \right) + \gamma^{(2)}_{n,0} \left( A_{n,0}^{(1)} g_{n,0} + A_{n,0} g'_{n,0} \right) \right).
\label{oneloopconformalblock}
  \end{aligned}
\end{equation}
Be aware here that $g^{(2)}(\nu_n)=\frac{1}{2}g''_{n,0}$. Plugging in (\ref{Dcross-2}) and (\ref{OPEcoeff1storder}), we found that two terms in (\ref{oneloopspecfunctionres}) each independently gives a consistent expression of the second order anomalous dimension, 
\begin{equation}
  \begin{aligned}
\gamma^{(2)}_{n,0} =& \frac{2\pi i}{A_{n,0}}D^{\times^{(-2)}}_{\nu_n} B^{(0)}_{\nu_n} =\gamma^{(1)}_{n,0}B^{(0)}_{\nu_n},
  \end{aligned}
\end{equation}
\begin{equation}
  \begin{aligned}
\gamma^{(2)}_{n,0} =& \frac{2\pi i}{A^{(1)}_{n,0}}D^{\times^{(-1)}}_{\nu_n} B^{(0)}_{\nu_n} =\gamma^{(1)}_{n,0}B^{(0)}_{\nu_n}.
  \end{aligned}
\end{equation}
Similarly, the second order OPE coefficient is found to be
\begin{equation}
  \begin{aligned}
A^{(2)}_{n,0} =& \frac{2\pi i}{(\gamma^{(1)}_{n,0})^2}2(D^{\times^{(0)}}_{\nu_n} B^{(-1)}_{\nu_n} + D^{\times^{(-2)}}_{\nu_n}B^{(1)})_{\nu_n}.
  \end{aligned}
\end{equation}
The first term in (\ref{oneloopspecfunctionres}) provides also a consistency check that
\begin{equation}
  \begin{aligned}
B^{(-1)}_{\nu_n} =\frac{1}{2\pi iD^{\times^{(-2)}}_{\nu_n}}(\gamma^{(1)}_{n,0})^2A_{n,0}=-\gamma^{(1)}_{n,0},
  \label{consistencycondition}
  \end{aligned}
\end{equation}
giving rise to a constraint on $B^{(-1)}_{\nu_n}$,

\begin{equation}
\resizebox{0.95\textwidth}{!}{$
  \begin{aligned}
B^{(-1)}_{\nu_n}=\frac{(-1)^n \pi ^{2-\frac{3 d}{2}} \lambda  \Gamma\left(\frac{d}{2}+n\right) \Gamma(d-n-2 \Delta ) \Gamma(\Delta )^2
\Gamma(n+\Delta )^2 \Gamma\left(-\frac{d}{2}+n+2 \Delta \right) \Gamma\left(\frac{d-2n-2\Delta}{2} \right)^2}{16 n! \Gamma\left(\frac{d}{2}\right)
\Gamma\left(\frac{d}{2}-\Delta \right)^2 \Gamma\left(1-\frac{d}{2}+\Delta \right)^4 \Gamma(2n+2\Delta )) \Gamma(d-2n-2\Delta
))}.
  \end{aligned}
$}
\end{equation}
Up to normalization, this is is in fact the same argument as \cite[4.24]{Carmi:2018qzm}, that $B^{(-1)}_{\nu_n}$ must have residues at $\frac{d}{2}+i\nu=2\Delta+2n$ which cancels precisely with the contribution from the generalized free field theory. It is one of the constraints that allowed for the bootstrap of an explicit expression of $B_{\nu_n}$. The first few OPE coefficients squared  at $\mathcal{O}(\lambda)$ are given in Table \ref{OPECoeff1stOrder}.

\begin{table}[h!]
\centering
\renewcommand{\arraystretch}{1.2}
\begin{tabular}{
| >{\centering\arraybackslash}m{1cm}
| >{\centering\arraybackslash}m{10cm} |
}
\hline
 & $A^{(1)}_{n,0}$ \\
\hline
$n=0$ &
$\tfrac{5}{2} - 4 \log 2$ \\
\hline
$n=1$ &
$\tfrac{69}{64} - \tfrac{9}{4}\log 2$ \\
\hline
$n=2$ &
$\tfrac{1755}{8192} - \tfrac{4050}{8192}\log 2$ \\
\hline
$n=3$ &
$\tfrac{48335}{1572864} - \tfrac{117600}{1572864}\log 2$ \\
\hline
$n=4$ &
$\tfrac{15789375}{4294967296} - \tfrac{39690000}{4294967296}\log 2$ \\
\hline
$n=5$ &
$\tfrac{336159}{85899345920}
- \tfrac{865125}{85899345920}\log 2$ \\
\hline
$n=6$ &
$\tfrac{13520857669}{351843720888320}
- \tfrac{17656192720}{351843720888320}\log 2$ \\
\hline
\end{tabular}
\caption{First few squared OPE coefficients at $\mathcal{O}(\lambda)$}
\label{OPECoeff1stOrder}
\end{table}

\section{Conformal Block}
\label{AppendixCBDef}
The conformal block with an intermediate operator dimension $\tilde \Delta=2\Delta+2n+l$ is given in \cite{Li:2019cwm} by
\begin{equation}
    G_{\tilde \Delta(n,l),l}=\sum\limits_{k=0}^\infty u^{\frac{\Delta-l}{2}+k}\sum\limits_{m=0}^{2k}A_{k,m}f_{k,m}(1-v),
\end{equation}
with
\begin{equation}
    \begin{aligned}
	    f_{k,m}(1-v)=&(1-v)^{l-m}
	    {}_2F_1\left({\tilde \Delta+l\over2}+k-m-a,{\tilde \Delta+l\over2}+k-m+b, \tilde \Delta+l+2k-2m;1-v\right) \\
        =&\sum_{p=0}^\infty \frac{\left({\tilde \Delta+l\over2}+k-m-a\right)_{p}^2} {p!\left(\tilde \Delta+l+2k-2m\right)_p}(1-v)^{l-m+p},
    \end{aligned}
\end{equation}
and
\begin{multline}
    A_{k,m}(\tilde \Delta)=\sum\limits_{m_1,m_2=0}^{\lfloor\frac{m}{2}\rfloor}\frac{1}{2^{l-1}}(-1)^{m+m_1+1}4^{m_1+m_2}
    \frac{(-l)_m(-\lfloor m/2\rfloor))_{m_1+m_2}(k-\lfloor m/2\rfloor)+1/2)_{m_1}}{m! m_1! m_2!(k-m+m_1)!}\cr
    \times\frac{(\tilde \Delta-1)_{2k-m}(d/2-\tilde \Delta)_{m-k-m_1-m_2}(l-\tilde \Delta+d-1)_{2(\lfloor m/2\rfloor-m_2)-m}}{(\tilde \Delta+l-m-1)_{2k-m}(\tilde \Delta+l)_{2(k+m1-\lfloor m/2\rfloor)-m}}\cr
    \times\frac{(1-d/2-l-k+m-m_1+m_2)\,(3/2-d/2-\ell+\lfloor (m+1)/2\rfloor)_{m_2}}
        {(2-d/2-l)_{-k+m+m_2}\,(-1+d/2+l-m_2)_{k-m+m_1+m_2+1}}\cr
    \times\prod\limits_{\alpha\in\{\pm a,\pm b\}}\left(\left(\frac12(\tilde \Delta+l)+\alpha\right)_{k-m+m_1}\left(\frac12(\tilde \Delta-l-d+2)+\alpha\right)_{m_2}\right)(1+(4ab-1)(m\text{ mod } 2)),
    \label{CBFactorExplicit}
\end{multline}
where $a=\frac{\Delta_1-\Delta_2}{2}$ and $b=\frac{\Delta_3-\Delta_4}{2}$. $(x)_n:=\frac{\Gamma(x+n)}{\Gamma(x)}$ is the Pochhammer symbol.

The squared OPE coefficients are given in \cite{Fitzpatrick:2011dm} by

{\footnotesize\begin{align}
    \label{OPECoefficientExplicit}
    A^{i,j}_{[\Op_i\Op_j]_{n,l}}&=\frac{2^l(-1)^l\left(\Delta_i-\frac{d}{2}+1\right)_n\left(\Delta_j-\frac{d}{2}+1\right)_n(\Delta_i)_{l+n}(\Delta_j)_{l+n}}
    {l!n!\left(l+\frac{d}{2}\right)_n(\Delta_i+\Delta_j+n-d+1)_n(\Delta_i+\Delta_j+2n+l-1)_l\left(\Delta_i+\Delta_j+n+l-\frac{d}{2}\right)_n}\,,
\end{align}}

Given the conformal block, it is also possible to determine the OPE coefficients from (\ref{CBExpansionFreeTheory}) and (\ref{FreeTheoryCorrelator}), yielding the same result as above. Note that $\frac{1}{2^{l-1}}$ and $2^l$ were added to (\ref{CBFactorExplicit}) and (\ref{OPECoefficientExplicit}) so that our normalization matches that in \cite{Heckelbacher:2022fzi}.

For arbitrary $n$ and $l$, the condition for $G_{n,l}$ to contribute a $u^i(1-v)^j$ is that

\begin{equation}
\begin{aligned}
\frac{\tilde \Delta-l}{2}=\frac{2\Delta+2n+l-l}{2}\leq \Delta+i,
\end{aligned}
\end{equation}

so that $n \leq i$ and $k=i-n$. In addition,

\begin{equation}
\begin{aligned}
l-m \leq l-2k \leq l-2i+2n \leq j.
\end{aligned}
\end{equation}

In the conformal block expansion (\ref{eq:CBExp}), contribution to the $u^i(1-v)^j$ term is therefore

\begin{equation}
\begin{aligned}
\sum_{n=0}^i\sum_{l=0,l\text{ even}}^{2i+j-2n}\mathcal{A}_{n,l} \mathcal{G}_{n,l}.
\end{aligned}
\end{equation}

\section{Recursive Relation}

\label{AppendixRecursiveRelation}

Any $\gamma^{(2),t+u}_{n,l}$ can be computed with a recursive relation and 5 initial values. The initial values are

\begin{equation}
  \begin{aligned}
\gamma^{(2),t+u}_{n,l} =
\begin{cases}
-\dfrac{\lambda^2}{96\pi^2}, & (n,l)=(0,0), \\[6pt]
-\dfrac{\lambda^2}{3360\pi^2}, & (n,l)=(0,2), \\[6pt]
-\dfrac{\lambda^2}{15840\pi^2}, & (n,l)=(0,4), \\[6pt]
-\dfrac{13\lambda^2}{3360\pi^2}, & (n,l)=(1,0), \\[6pt]
-\dfrac{61\lambda^2}{221760\pi^2}, & (n,l)=(1,2).
\end{cases}
  \end{aligned}
\end{equation}

\begin{figure}[t]
\centering
\resizebox{0.4\linewidth}{!}{%
\begin{tikzpicture}
  \begin{feynman}
    \vertex (nltext) at (-1,1) {\(\gamma^{(2),t+u}_{n,l}\)};
    \vertex (l1) at (0,1);
    \vertex (l2) at (3,1);
    \vertex [above=0.3cm of l2] (l2text) {\(l\)};
    \vertex (n1) at (-1,0);
    \vertex (n2) at (-1,-3);
    \vertex [left=0.3cm of n2] (n2text) {\(n\)};
    \node [dot] (1-0) at (1,0);
    \node [dot] (2-0) at (2,0);
    \node [dot] (1-1) at (1,-1);
    \node [dot] (0-1) at (0,-1);
    \node [dot, green] (3-0) at (3,0);
    \node [dot, green] (5-4) at (2,-1);
    \node [dot, green] (5-5) at (1,-2);
    \node [dot, green] (4-5) at (0,-2);

    \draw[->, thick] ($ (l1)!0.0!(l2) $) to ($ (l1)!1.0!(l2) $);
    \draw[->, thick] ($ (n1)!0.0!(n2) $) to ($ (n1)!1.0!(n2) $);
  \end{feynman}
\end{tikzpicture}%
}
\caption{Recursive relations}
\label{RecursiveRelationDotFigure}
\end{figure}
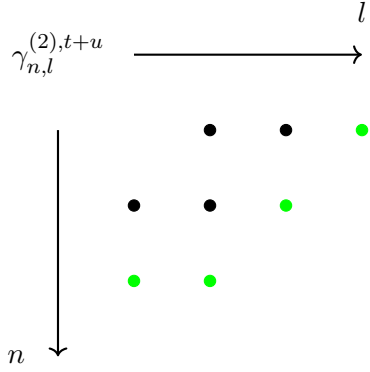

For any $n \geq 1$ and $l \geq 2$, $\gamma^{(2),t+u}_{n+1,l-2}$, $\gamma^{(2),t+u}_{n+1,l}$, $\gamma^{(2),t+u}_{n,l+2}$, $\gamma^{(2),t+u}_{n-1,l+4}$ (marked in Figure \ref{RecursiveRelationDotFigure} as green dots) can be expressed as polynomials of $n$ and $l$, together with $\gamma^{(2),t+u}_{n,l}$, $\gamma^{(2),t+u}_{n-1,l}$, $\gamma^{(2),t+u}_{n-1,l+2}$ (marked as black dots). The explicit expressions of these polynomials are given in (\ref{RecursiveRelationPolynomial1})-(\ref{RecursiveRelationPolynomial4}). This recursive relation is also implemented as a Mathematica notebook, which is available \href{https://github.com/weichenxiao/The-2-Dimensional-Dual-of-phi4-in-AdS3}{here}.

\begin{equation}
\begin{aligned}
&\text{When $n \geq 2$,} \\
&\gamma^{(2),t+u}_{n,l}=\Bigl(
- (1 + l + n) \bigl(
55 + 8 l^{3} + 126 n + 64 n^{2} + 4 l^{2}(13 + 6 n) + 2 l(51 + 60 n + 8 n^{2})
\bigr) \gamma^{(2),t+u}_{n-2,\,l+2} \\
&\quad + (35 + 24 l + 4 l^{2}) (3 + l + n) (5 + 2 l + 2 n) \gamma^{(2),t+u}_{n-2,\,l+4} \\
&\quad - (2 + l + n) \bigl(
15 + 8 l^{3} + l^{2}(36 - 8 n) - 114 n - 64 n^{2} - 2 l(-23 + 36 n + 8 n^{2})
\bigr) \gamma^{(2),t+u}_{n-1,\,l+2} \\
&\quad + \bigl(
8 l^{4} + 4 l^{3}(9 + 8 n) + 2 l^{2}(23 + 54 n + 20 n^{2}) + n(15 + 46 n + 32 n^{2}) \\
&\quad\quad + l(15 + 92 n + 104 n^{2} + 16 n^{3})
\bigr) \gamma^{(2),t+u}_{n-1,\,l}
\Bigr) \\
&\times \Bigl(
8 (1 + n)(1 + l + n)(5 + 2 l^{2} + 4 n + l (7 + 2 n))
\Bigr)^{-1}.
\end{aligned}
\label{RecursiveRelationPolynomial1}
\end{equation}

\newpage

\begin{equation}
\begin{aligned}
&\text{when $n \geq 1$ and $l \geq 4$.} \\
&\gamma^{(2),t+u}_{n,l}=\Bigl(
-2 n \bigl(
166 + 16 l^{5} + 13 n - 632 n^{2} + 548 n^{3} - 128 n^{4}
+ 16 l^{4}(-8 + 5 n) \\
&\quad + 8 l^{3}(41 - 72 n + 18 n^{2})
+ 8 l^{2}(-31 + 159 n - 112 n^{2} + 14 n^{3}) \\
&\quad + l(-131 - 888 n + 1492 n^{2} - 576 n^{3} + 32 n^{4})
\bigr) \gamma^{(2),t+u}_{n-1,\,l-2} \\
&\quad + 2 n \bigl(
16 l^{5} + 16 l^{4}(-2 + 5 n)
+ 8 l^{3}(-7 - 24 n + 18 n^{2}) \\
&\quad\quad + n(-3 + 64 n + 36 n^{2} - 64 n^{3})
+ 8 l^{2}(9 - 9 n - 44 n^{2} + 14 n^{3}) \\
&\quad\quad + l(-3 + 136 n + 20 n^{2} - 256 n^{3} + 32 n^{4})
\bigr) \gamma^{(2),t+u}_{n-1,\,l} \\
&\quad + \bigl(
525 - 220 l - 1401 l^{2} + 1792 l^{3} - 872 l^{4} + 192 l^{5} - 16 l^{6} \\
&\quad\quad + 665 n - 4621 l n + 6292 l^{2} n - 3480 l^{3} n + 864 l^{4} n - 80 l^{5} n \\
&\quad\quad - 2800 n^{2} + 6120 l n^{2} - 4460 l^{2} n^{2} + 1344 l^{3} n^{2} - 144 l^{4} n^{2} \\
&\quad\quad + 1680 n^{3} - 2132 l n^{3} + 864 l^{2} n^{3} - 112 l^{3} n^{3} \\
&\quad\quad - 280 n^{4} + 192 l n^{4} - 32 l^{2} n^{4}
\bigr) \gamma^{(2),t+u}_{n,\,l-4} \\
&\quad + \bigl(
-30 + 92 l + 18 l^{2} - 320 l^{3} + 400 l^{4} - 192 l^{5} + 32 l^{6} \\
&\quad\quad + 36 n + 468 l n - 1816 l^{2} n + 2240 l^{3} n - 1120 l^{4} n + 192 l^{5} n \\
&\quad\quad + 394 n^{2} - 2496 l n^{2} + 4104 l^{2} n^{2} - 2432 l^{3} n^{2} + 448 l^{4} n^{2} \\
&\quad\quad - 1032 n^{3} + 3152 l n^{3} - 2528 l^{2} n^{3} + 512 l^{3} n^{3} \\
&\quad\quad + 888 n^{4} - 1280 l n^{4} + 288 l^{2} n^{4} - 256 n^{5} + 64 l n^{5}
\bigr) \gamma^{(2),t+u}_{n,\,l-2}
\Bigr) \\
&\times \Bigl(
15 + 16 l^{6} + 113 n + 112 l^{5} n + 250 n^{2} + 120 n^{3} - 160 n^{4} - 128 n^{5} \\
&\quad + 8 l^{4}(-11 - 8 n + 38 n^{2})
+ 8 l^{3}(-8 - 59 n - 40 n^{2} + 50 n^{3}) \\
&\quad + l^{2}(57 - 132 n - 868 n^{2} - 576 n^{3} + 256 n^{4}) \\
&\quad + l(64 + 263 n + 40 n^{2} - 644 n^{3} - 448 n^{4} + 64 n^{5})
\Bigr)^{-1}.
\end{aligned}
\label{RecursiveRelationPolynomial2}
\end{equation}

\newpage

\begin{equation}
\begin{aligned}
&\text{When $n=0$ and $l \geq 6$,} \\
&\gamma^{(2),t+u}_{n,l}=\Biggl(
- \bigl(
-12339 + 128 l^9 - 8361 n + 176934 n^2 \\
&\quad  + 25280 n^3 - 111672 n^4 + 108656 n^5 - 51456 n^6 + 8192 n^7 \\
&\quad + 64 l^8 (-27 + 20 n) + 32 l^7 (241 - 558 n + 152 n^2) + 144 l^6 (-35 + 644 n - 460 n^2 + 64 n^3) \\
&\quad + 8 l^5 (-7313 - 23706 n + 43708 n^2 - 14760 n^3 + 1168 n^4) \\
&\quad + 4 l^4 (42747 - 520 n - 204336 n^2 + 149568 n^3 - 27360 n^4 + 1216 n^5) \\
&\quad + 2 l^3 (-77493 + 265566 n + 343520 n^2 - 665136 n^3 + 261856 n^4 - 25344 n^5 + 512 n^6) \\
&\quad + l (42282 + 253365 n - 423578 n^2 - 292884 n^3 + 657736 n^4 - 352896 n^5 + 60160 n^6)
\bigr) \gamma^{(2),t+u}_{n,\,l-4} \\
&\quad + 2 \bigl(
-387 + 128 l^9 + 72 n + 19417 n^2 - 17754 n^3 - 26204 n^4 + 34584 n^5 - 9728 n^6 \\
&\quad + 64 l^8 (-19 + 18 n) + 32 l^7 (121 - 348 n + 128 n^2) + 16 l^6 (-251 + 2366 n - 2456 n^2 + 464 n^3) \\
&\quad + 8 l^5 (-401 - 6312 n + 17076 n^2 - 8608 n^3 + 912 n^4) \\
&\quad + 4 l^4 (2467 - 402 n - 52444 n^2 + 58088 n^3 - 15984 n^4 + 928 n^5) \\
&\quad + 2 l^3 (-3021 + 36196 n + 53480 n^2 - 179056 n^3 + 99904 n^4 - 15040 n^5 + 384 n^6) \\
&\quad + 2 l (825 + 10472 n - 39751 n^2 - 13724 n^3 + 80620 n^4 - 47184 n^5 + 6432 n^6)
\bigr) \gamma^{(2),t+u}_{n,\,l-2} \\
&\quad + (2+n) \biggl(
(99 - 40 l + 4 l^2) \bigl(
-84 + 32 l^6 + 45 n + 1002 n^2 - 540 n^3 + 72 n^4 \\
&\quad + 16 l^5 (-19 + 12 n) + 16 l^4 (59 - 95 n + 28 n^2) + 8 l^3 (-125 + 480 n - 342 n^2 + 64 n^3) \\
&\quad + l^2 (98 - 3496 n + 4920 n^2 - 2128 n^3 + 288 n^4) \\
&\quad + l (269 + 1212 n - 3036 n^2 + 2096 n^3 - 608 n^4 + 64 n^5)
\bigr) \gamma^{(2),t+u}_{n+1,\,l-6} \\
&\quad - \bigl(
-990 + 128 l^8 + 1203 n + 30790 n^2 - 32260 n^3 - 8808 n^4 + 16768 n^5 - 4096 n^6 \\
&\quad + 192 l^7 (-9 + 4 n) + 32 l^6 (273 - 310 n + 56 n^2) + 16 l^5 (-1335 + 2908 n - 1340 n^2 + 128 n^3) \\
&\quad + 8 l^4 (3231 - 13574 n + 10500 n^2 - 2760 n^3 + 144 n^4) \\
&\quad + 4 l^3 (-3009 + 35180 n - 40240 n^2 + 14608 n^3 - 2720 n^4 + 64 n^5) \\
&\quad - 2 l^2 (1525 + 50346 n - 90744 n^2 + 33392 n^3 - 1600 n^4 + 1024 n^5) \\
&\quad + l (4371 + 30636 n - 119764 n^2 + 58032 n^3 + 23712 n^4 - 13120 n^5)
\bigr) \gamma^{(2),t+u}_{n+1,\,l-4}
\biggr)
\Biggr) \\
&\times \Bigl(
(3 - 8 l + 4 l^2) (1 + l + n) (1 + 2 l + 2 n) \\
&\quad \times \bigl(
-57 + 16 l^5 - 166 n + 384 n^2 + 480 n^3 - 256 n^4 + 16 l^4 (-5 + 6 n) \\
&\quad + 8 l^3 (-1 - 62 n + 26 n^2) + 8 l^2 (31 + 28 n - 132 n^2 + 24 n^3) \\
&\quad + l (1 + 748 n + 732 n^2 - 896 n^3 + 64 n^4)
\bigr)
\Bigr)^{-1}.
\end{aligned}
\label{RecursiveRelationPolynomial3}
\end{equation}

\newpage

\begin{equation}
\begin{aligned}
&\text{When $n \geq 2$ and $l\geq2$,} \\
&\gamma^{(2),t+u}_{n,l}=\Bigl(
(3 + 8 l + 4 l^{2}) \bigl(
-5 + 8 l^{4} - 7 n - 2 n^{2} + 4 l^{3}(3 + 8 n) \\
&\quad + l^{2}(-6 + 20 n + 40 n^{2}) + l(-15 - 20 n + 8 n^{2} + 16 n^{3})
\bigr) \gamma^{(2),t+u}_{n-2,\,l+2} \\
&\quad + \bigl(
-39 - 32 l^{6} + l^{5}(16 - 192 n) + 89 n + 46 n^{2} - 224 n^{3} + 128 n^{4} \\
&\quad + l^{4}(224 + 80 n - 544 n^{2}) - 8 l^{3}(45 - 112 n - 40 n^{2} + 112 n^{3}) \\
&\quad - 2 l^{2}(-51 + 540 n - 584 n^{2} - 320 n^{3} + 384 n^{4}) \\
&\quad + l(89 + 204 n - 944 n^{2} + 544 n^{3} + 384 n^{4} - 256 n^{5})
\bigr) \gamma^{(2),t+u}_{n-2,\,l} \\
&\quad + \bigl(
-18 + 69 l - 14 l^{2} - 232 l^{3} + 336 l^{4} - 176 l^{5} + 32 l^{6} \\
&\quad + 21 n + 76 l n - 576 l^{2} n + 944 l^{3} n - 592 l^{4} n + 128 l^{5} n \\
&\quad - 6 n^{2} - 152 l n^{2} + 560 l^{2} n^{2} - 544 l^{3} n^{2} + 160 l^{4} n^{2} \\
&\quad + 48 l n^{3} - 128 l^{2} n^{3} + 64 l^{3} n^{3}
\bigr) \gamma^{(2),t+u}_{n-1,\,l-2} \\
&\quad + \bigl(
-9 l + 30 l^{2} + 8 l^{3} - 112 l^{4} + 112 l^{5} - 32 l^{6} \\
&\quad - 9 n + 36 l n - 144 l^{2} n - 656 l^{3} n + 720 l^{4} n \\
&\quad + 6 n^{2} - 312 l n^{2} - 1264 l^{2} n^{2} + 1376 l^{3} n^{2} + 608 l^{4} n^{2} \\
&\quad - 160 n^{3} - 976 l n^{3} + 1024 l^{2} n^{3} + 1600 l^{3} n^{3} \\
&\quad - 256 n^{4} + 256 l n^{4} + 1536 l^{2} n^{4} + 512 l n^{5}
\bigr) \gamma^{(2),t+u}_{n-1,\,l}
\Bigr) \\
&\times \Bigl(
8 (1+n)(1+l+n) \bigl(
-3 + 8 l^{4} - 16 n - 16 n^{2} + 4 l^{3}(3 + 10 n) \\
&\quad + l^{2}(-6 + 32 n + 64 n^{2}) + l(-11 - 26 n + 16 n^{2} + 32 n^{3})
\bigr)
\Bigr)^{-1}.
\label{RecursiveRelationPolynomial4}
\end{aligned}
\end{equation}




\bibliographystyle{JHEP}
\bibliography{megabib}

\end{document}